\documentclass[preprint2]{aastex62}
\usepackage{amsmath,amssymb,comment,gensymb,multirow,soul}
\RequirePackage{lineno}
\received{2021 June 15}
\revised{2021 October 20}
\accepted{2021 October 22}
\submitjournal{The Astrophysical Journal}

\shorttitle{Three Gamma-ray Quasars Observed by VERITAS and \textit{Fermi}-LAT}
\shortauthors{VERITAS et al.}
\begin{document}
%\linenumbers
\title{Variability and Spectral Characteristics of Three Flaring Gamma-ray Quasars Observed by VERITAS and \textit{Fermi}-LAT}

\correspondingauthor{Aryeh Brill}
\email{aryeh.brill@nasa.gov}

\correspondingauthor{Olivier Hervet}
\email{ohervet@ucsc.edu}

\correspondingauthor{Janeth Valverde}
\email{valverde@llr.in2p3.fr}

\author{C.~B.~Adams}
\affiliation{Department of Physics and Astronomy, Barnard College, Columbia University, NY 10027, USA}

\author{J.~Batshoun}
\affiliation{Department of Physics, California State University - East Bay, Hayward, CA 94542, USA} % Undergrad

\author{W.~Benbow}
\affiliation{Center for Astrophysics $|$ Harvard \& Smithsonian, Cambridge, MA 02138, USA}

\author{A.~Brill}
\affiliation{Physics Department, Columbia University, New York, NY 10027, USA}
\affiliation{Now at NASA Goddard Space Flight Center, Greenbelt, MD 20771, USA}

\author{J.~H.~Buckley}
\affiliation{Department of Physics, Washington University, St. Louis, MO 63130, USA}

\author{M.~Capasso}
\affiliation{Department of Physics and Astronomy, Barnard College, Columbia University, NY 10027, USA}

\author{B.~Cavins}
\affiliation{Santa Cruz Institute for Particle Physics and Department of Physics, University of California, Santa Cruz, CA 95064, USA}

\author{J.~L.~Christiansen}
\affiliation{Physics Department, California Polytechnic State University, San Luis Obispo, CA 94307, USA}

\author{P.~Coppi}\affiliation{260 Whitney Avenue, Yale University, New Haven, CT 06511, USA}

\author{M.~Errando}
\affiliation{Department of Physics, Washington University, St. Louis, MO 63130, USA}

\author{K.~A~Farrell}
\affiliation{School of Physics, University College Dublin, Belfield, Dublin 4, Ireland}

\author{Q.~Feng}
\affiliation{Department of Physics and Astronomy, Barnard College, Columbia University, NY 10027, USA}

\author{J.~P.~Finley}
\affiliation{Department of Physics and Astronomy, Purdue University, West Lafayette, IN 47907, USA}

\author{G.~M.~Foote}
\affiliation{Department of Physics and Astronomy and the Bartol Research Institute, University of Delaware, Newark, DE 19716, USA}

\author{L.~Fortson}
\affiliation{School of Physics and Astronomy, University of Minnesota, Minneapolis, MN 55455, USA}

\author{A.~Furniss}
\affiliation{Department of Physics, California State University - East Bay, Hayward, CA 94542, USA}

\author{A.~Gent}
\affiliation{School of Physics and Center for Relativistic Astrophysics, Georgia Institute of Technology, 837 State Street NW, Atlanta, GA 30332-0430}

\author{C.~Giuri}
\affiliation{DESY, Platanenallee 6, 15738 Zeuthen, Germany}

\author{D.~Hanna}
\affiliation{Physics Department, McGill University, Montreal, QC H3A 2T8, Canada}

\author{T.~Hassan}
\affiliation{DESY, Platanenallee 6, 15738 Zeuthen, Germany}

\author{O.~Hervet}
\affiliation{Santa Cruz Institute for Particle Physics and Department of Physics, University of California, Santa Cruz, CA 95064, USA}

\author{J.~Holder}
\affiliation{Department of Physics and Astronomy and the Bartol Research Institute, University of Delaware, Newark, DE 19716, USA}

\author{M.~Houck}
\affiliation{Davidson College, 405 N Main St, Davidson, NC 28035, USA}
\affiliation{Department of Civil and Environmental Engineering, University of California, Irvine, CA 92617, USA} % Undergrad

\author{T.~B.~Humensky}
\affiliation{Physics Department, Columbia University, New York, NY 10027, USA}

\author{W.~Jin}
\affiliation{Department of Physics and Astronomy, University of Alabama, Tuscaloosa, AL 35487, USA}

\author{P.~Kaaret}
\affiliation{Department of Physics and Astronomy, University of Iowa, Van Allen Hall, Iowa City, IA 52242, USA}

\author{M.~Kertzman}
\affiliation{Department of Physics and Astronomy, DePauw University, Greencastle, IN 46135-0037, USA}

\author{D.~Kieda}
\affiliation{Department of Physics and Astronomy, University of Utah, Salt Lake City, UT 84112, USA}

\author{F.~Krennrich}
\affiliation{Department of Physics and Astronomy, Iowa State University, Ames, IA 50011, USA}

\author{S.~Kumar}
\affiliation{Physics Department, McGill University, Montreal, QC H3A 2T8, Canada}

\author{M.~Lundy}
\affiliation{Physics Department, McGill University, Montreal, QC H3A 2T8, Canada}

\author{G.~Maier}
\affiliation{DESY, Platanenallee 6, 15738 Zeuthen, Germany}

\author{C.~E~McGrath}
\affiliation{School of Physics, University College Dublin, Belfield, Dublin 4, Ireland}

\author{P.~Moriarty}
\affiliation{School of Physics, National University of Ireland Galway, University Road, Galway, Ireland}

\author{R.~Mukherjee}
\affiliation{Department of Physics and Astronomy, Barnard College, Columbia University, NY 10027, USA}

\author{D.~Nieto}
\affiliation{Institute of Particle and Cosmos Physics, Universidad Complutense de Madrid, 28040 Madrid, Spain}

\author{M.~Nievas-Rosillo}
\affiliation{DESY, Platanenallee 6, 15738 Zeuthen, Germany}

\author{S.~O'Brien}
\affiliation{Physics Department, McGill University, Montreal, QC H3A 2T8, Canada}

\author{R.~A.~Ong}
\affiliation{Department of Physics and Astronomy, University of California, Los Angeles, CA 90095, USA}

\author{A.~Oppenheimer}
\affiliation{Hastings High School, 1 Mt Hope Blvd, Hastings-On-Hudson, NY 10706, USA}
\affiliation{University of Chicago, Chicago, IL 60637, USA} % High school student

\author{A.~N.~Otte}
\affiliation{School of Physics and Center for Relativistic Astrophysics, Georgia Institute of Technology, 837 State Street NW, Atlanta, GA 30332-0430}

\author{S.~Patel}
\affiliation{Department of Physics and Astronomy, University of Iowa, Van Allen Hall, Iowa City, IA 52242, USA}

\author{K.~Pfrang}
\affiliation{DESY, Platanenallee 6, 15738 Zeuthen, Germany}

\author{M.~Pohl}
\affiliation{Institute of Physics and Astronomy, University of Potsdam, 14476 Potsdam-Golm, Germany and DESY, Platanenallee 6, 15738 Zeuthen, Germany}

\author{R.~R.~Prado}
\affiliation{DESY, Platanenallee 6, 15738 Zeuthen, Germany}

\author{E.~Pueschel}
\affiliation{DESY, Platanenallee 6, 15738 Zeuthen, Germany}

\author{J.~Quinn}
\affiliation{School of Physics, University College Dublin, Belfield, Dublin 4, Ireland}

\author{K.~Ragan}
\affiliation{Physics Department, McGill University, Montreal, QC H3A 2T8, Canada}

\author{P.~T.~Reynolds}
\affiliation{Department of Physical Sciences, Munster Technological University, Bishopstown, Cork, T12 P928, Ireland}

\author{A.~Rhatigan}
\affiliation{Physics Department, California Polytechnic State University, San Luis Obispo, CA 94307, USA} % Undergrad

\author{D.~Ribeiro}
\affiliation{Physics Department, Columbia University, New York, NY 10027, USA}

\author{E.~Roache}
\affiliation{Center for Astrophysics $|$ Harvard \& Smithsonian, Cambridge, MA 02138, USA}

\author{J.~L.~Ryan}
\affiliation{Department of Physics and Astronomy, University of California, Los Angeles, CA 90095, USA}

\author{M.~Santander}
\affiliation{Department of Physics and Astronomy, University of Alabama, Tuscaloosa, AL 35487, USA}

\author{G.~H.~Sembroski}
\affiliation{Department of Physics and Astronomy, Purdue University, West Lafayette, IN 47907, USA}

\author{D.~A.~Williams
}\affiliation{Santa Cruz Institute for Particle Physics and Department of Physics, University of California, Santa Cruz, CA 95064, USA}

\author{T.~J~Williamson}
\affiliation{Department of Physics and Astronomy and the Bartol Research Institute, University of Delaware, Newark, DE 19716, USA}

\collaboration{(VERITAS Collaboration)}

\author{J.~Valverde}
\affiliation{Center for Space Sciences and Technology, University of Maryland Baltimore County, Baltimore, MD 21250, USA}
\affiliation{NASA Goddard Space Flight Center, Greenbelt, MD 20771, USA}
\affiliation{Laboratoire Leprince-Ringuet, \'Ecole Polytechnique, CNRS/IN2P3, 91128 Palaiseau, France}
\affiliation{Department of Astronomy, Columbia University, New York, NY 10027, USA}

\author{D.~Horan}
\affiliation{Laboratoire Leprince-Ringuet, \'Ecole Polytechnique, CNRS/IN2P3, 91128 Palaiseau, France}

\author{S.~Buson}
\affiliation{Julius-Maximilians-Universit\"{a}t, 97070, W\"{u}rzburg, Germany}

\author{C.~C.~Cheung}
\affiliation{Space Science Division, Naval Research Laboratory, Washington, DC 20375, USA}

\author{S.~Ciprini}
\affiliation{Istituto Nazionale di Fisica Nucleare (INFN) Sezione di Roma Tor Vergata, Via della Ricerca Scientica 1, 00133, Roma, Italy}
\affiliation{ASI Space Science Data Center (SSDC), Via del Politecnico, 00133, Roma, Italy}

\author{D.~Gasparrini}
\affiliation{Istituto Nazionale di Fisica Nucleare, Sezione di Roma “Tor Vergata”, I-00133 Roma, Italy}
\affiliation{Space Science Data Center - Agenzia Spaziale Italiana, Via del Politecnico, snc, I-00133, Roma, Italy}

\author{R. Ojha}
\affiliation{NASA HQ, Washington, DC 20546, USA}

\author{P.~van~Zyl}
\affiliation{Hartebeesthoek Radio Astronomy Observatory (HartRAO)}
\affiliation{South African Radio Astronomy Observatory (SARAO)}

\collaboration{(Fermi-LAT Collaboration)}

\author{L.~Sironi}
\affiliation{Department of Astronomy and Columbia Astrophysics Laboratory, Columbia University, New York, NY 10027, USA}

\begin{abstract}

\noindent Flat spectrum radio quasars (FSRQs) are the most luminous blazars at GeV energies, but only rarely emit detectable fluxes of TeV gamma rays, typically during bright GeV flares. We explore the gamma-ray variability and spectral characteristics of three FSRQs that have been observed at GeV and TeV energies by \textit{Fermi}-LAT and VERITAS, making use of almost 100 hours of VERITAS observations spread over 10 years: 3C~279, PKS~1222+216, and Ton~599. We explain the GeV flux distributions of the sources in terms of a model derived from a stochastic differential equation describing fluctuations in the magnetic field in the accretion disk, and estimate the timescales of magnetic flux accumulation and stochastic instabilities in their accretion disks. We identify distinct flares using a procedure based on Bayesian blocks and analyze their daily and sub-daily variability and gamma-ray energy spectra. Using observations from VERITAS as well as \textit{Fermi}, \textit{Swift}, and the Steward Observatory, we model the broadband spectral energy distributions of PKS~1222+216 and Ton~599 during VHE-detected flares in 2014 and 2017, respectively, strongly constraining the jet Doppler factors and gamma-ray emission region locations during these events. Finally, we place theoretical constraints on the potential production of PeV-scale neutrinos during these VHE flares.

\end{abstract}


\section{Introduction}\label{sec:intro}

Blazars are a class of active galactic nuclei (AGN) with jets oriented nearly along our line of sight. This alignment produces beamed emission, so that many blazars  show superluminal motion in their jets \citep[e.g.][]{Jorstad2001} and have a gamma-ray luminosity dominating their bolometric power. In jet models, high-energy electrons in a relativistically outflowing jet, ejected from an accreting supermassive black hole, are responsible for the synchrotron radiation seen as the radio to UV continuum from blazars \citep{1979ApJ...232...34B}. Blazars have a spectral energy distribution (SED) exhibiting a double-humped structure, with low-energy synchrotron and high-energy gamma-ray-peaked components.

Blazars are the most common gamma-ray-emitting objects in the extragalactic sky. Observationally, they can be divided into two classes: BL Lacertae (BL Lac) objects, the aligned counterparts to Fanaroff-Riley I radio galaxies, and flat spectrum radio quasars (FSRQs), the counterparts to Fanaroff-Riley II radio galaxies \citep{Fanaroff1974}.
%While many BL Lac objects are high-synchrotron-peaked (HSP) blazars with the frequency of the synchrotron peak $\nu_S > 10^{15}$ Hz,
FSRQs are low-synchrotron-peaked (LSP) blazars, with synchrotron peak frequency less than $10^{14}$ Hz. The bolometric luminosity of FSRQs is typically greater than that of BL Lac objects. The anti-correlation of synchrotron luminosity with peak frequency is an empirical relationship known as the blazar sequence \citep{Fossati1998, Nieppola2008}, though its intrinsic validity has been disfavored by more recent work \citep{Keenan2021}. Accordingly, while FSRQs make up only 8 of the 79 AGN that have been detected in the TeV band to date\footnote{\url{http://tevcat.in2p3.fr/}}, they are more commonly detected at GeV energies, comprising 650 of 2863 AGN detected by the Large Area Telescope (LAT) on board the \textit{Fermi} Gamma-Ray space telescope (\textit{Fermi}-LAT) \citep{Ajello2020} and dominating the blazar population detected by the Energetic Gamma Ray Experiment Telescope (EGRET) on the Compton Gamma-Ray Observatory \citep{Mukherjee2001}.

%FSRQs can be identified by broad optical spectral lines ($>5\text{\AA}$) and flat radio spectra, with spectral index $\alpha \approx 0$.
The SED of an FSRQ is generally dominated by the gamma-ray emission component, which peaks in the high-energy (HE;~${\sim}$GeV) band. FSRQs are believed to possess several structures producing radiation fields external to the jet, including a broad line region (BLR) and a dust torus. TeV detections of FSRQs are particularly interesting because the external radiation fields might be expected to produce increased Compton cooling of electrons and to absorb energetic gamma rays by pair production, leading to a cutoff in the gamma-ray spectrum above the GeV band \citep[e.g.][]{Ghisellini1998}.

Blazars have been observed to be variable at all wavelengths and at timescales down to several minutes in both the GeV and TeV bands \citep{Ackermann2016, Aharonian2007}. However, the physical mechanisms that drive this variability are unclear. Different processes, possibly originating at different locations in the AGN, may drive variable emission occurring at different timescales. By providing an upper bound on the light crossing time, the timescale of variability constrains the apparent size of the emission region, giving information on the location and mechanism of the gamma-ray emission. While short variability timescales observed in blazars suggest that the emission may be connected to processes in the central engine or accretion disk, the ability of very-high-energy (VHE;~${\gtrsim}100$ GeV) emission to escape the AGN implies an origin further out in the jet, where absorption is reduced \citep{Abeysekara2015}.

Over longer timescales, blazar variability can be studied through the flux distribution describing the relative frequencies of different flux levels. Blazar flux distributions exhibit long tails, and have been described using log-normal models \citep[e.g.][]{Giebels09}, which could indicate evidence of an underlying multiplicative physical process. \citet{Meyer2019} fit the flux distributions of six bright FSRQs with a broken power law, though a log-normal distribution was also compatible with their data, and recently, \citet{Tavecchio2020} have described the gamma-ray flux variability of those same objects using a model based on a stochastic differential equation (SDE) including both deterministic and stochastic components.

The physical structure and multiwavelength emission mechanisms of a blazar can be further understood by modeling its SED.
%The low-energy SED component can be attributed to synchrotron radiation emitted by a population of relativistic electrons.
In leptonic models, the gamma-ray SED component is explained by relativistic electrons and positrons scattering via the inverse Compton process off of a population of lower-energy seed photons, which may be their own emitted synchrotron photons, as in the synchrotron self-Compton process \citep[SSC;][]{1992ApJ...397L...5M}, or radiation from an external structure, as in the external inverse Compton process \citep[EIC; e.g.][]{1996MNRAS.280...67G}. The EIC seed photons are commonly taken to be radiation fields in the BLR, although this picture has been challenged by the lack of characteristic BLR absorption features in the average gamma-ray spectra of \textit{Fermi}-LAT FSRQs \citep{Costamante2018}.

In hadronic models, however, some or all of the gamma-ray emission is due to relativistic protons emitting via photohadronic processes, proton synchrotron radiation, or other mechanisms, so that relativistic neutrino emission may occur as well. For example, the blazar high-energy emission may be dominated by synchrotron radiation losses of high-energy protons \citep[see e.g.][]{2000NewA....5..377A, 2000AIPC..515..149M}. Alternatively, neutrinos may be produced by the photohadronic interaction of a proton with a photon, producing pions that quickly decay to gamma rays and neutrinos, that is, $p\gamma \rightarrow \Delta^+ \rightarrow p\pi^0~\mathrm{or}~n\pi^+$ \citep{Dermer2009}. In this case, production of PeV-scale neutrinos requires a target photon population in the X-ray band. The $p\gamma$ process may co-occur with leptonic gamma-ray emission. Under this scenario, FSRQs may be sources of relativistic neutrinos at PeV or even EeV energies \citep[e.g.][]{Gao2017, Righi2020}. High-energy neutrinos have been detected coming from a direction compatible with the blazar TXS~0506+056 \citep{Aartsen2018}, which may be an FSRQ masquerading as a BL Lac object \citep{Padovani2019}.

% The first blazar associated with the production of high-energy neutrinos, TXS~0506+056 \citep{Aartsen2018}, may be an FSRQ masquerading as a BL Lac object \citep{Padovani2019}.

In this paper, we investigate strong gamma-ray flares from three FSRQs at intermediate redshifts. These three sources were continuously monitored by \textit{Fermi}-LAT (Section~\ref{sec:fermi}) during the ten-year period from 2008 to 2018, and observed during periods of high gamma-ray activity by the Very Energetic Radiation Imaging Telescope Array System (VERITAS;  Section~\ref{sec:VERITAS}). Table~\ref{tab:dataset} provides an overview of the gamma-ray data analyzed in this work.

3C~279, at $z=0.536$ \citep{Lynds1965}, is one of the most well-studied blazars. It is among the brightest and most variable extragalactic objects in the gamma-ray sky, giving rise to one of the first large amplitude gamma-ray flares measured by EGRET in 1996 \citep{Wehrle_1998}. In recent times, it underwent multiple bright gamma-ray flares in 2014, 2015, and 2018. Notably, during a flare beginning on June 16, 2015, it was detected by High Energy Stereoscopic System (H.E.S.S.), and \textit{Fermi}-LAT observed minute-scale variability \citep{Romoli2017, Ackermann2016}. H.E.S.S. again detected 3C~279 during the flaring states in January and June 2018 \citep{Emery2019}.

PKS~1222+216, at $z=0.432$ \citep{Osterbrock1987} and also known as 4C~+21.35, has exhibited periods of extreme variability in the VHE gamma-ray band, with VHE detections occurring during gamma-ray flares in June 2010 \citep{Aleksic2011} and February and March 2014 \citep{2014ATel.5981....1H}.

Finally, Ton~599, at $z=0.725$ (\citealp{Schneider2010}; see also \citealp{Burbidge1968}) and also known as 4C~+29.45 and B1156+295, entered a months-long GeV high state in October 2017 \citep{Cheung2017}, leading to VHE detections on the nights of December 15 and 16 2017 \citep{2017ATel11061....1M, 2017ATel11075....1M}.

We describe the observations and data analysis of these sources in Section~\ref{sec:analysis}. In Section~\ref{sec:fluxdistributions}, we examine the long-term variability of these FSRQs and connect it to processes in the accretion disk. Next, we select gamma-ray flares (Section~\ref{sec:flarealgorithm}) and analyze the short-timescale variability (Section~\ref{sec:flare_profiles}) and spectra (Section~\ref{sec:spectra}) during these events, focusing primarily on 3C~279, the brightest of the three sources in \textit{Fermi}-LAT. In Section~\ref{sec:sed_modeling}, we model the SEDs of PKS~1222+216 and Ton~599 during their respective VHE detections by VERITAS. The observed VHE emission places constraints on the Doppler factor and gamma-ray emission region location during these flares, which we confirm using an independent method in Section~\ref{sec:discussion_emission_region}. In Section~\ref{sec:discussion_neutrinos}, we place theoretical constraints on the potential production of PeV-scale neutrinos during these VHE flares. We summarize our conclusions in Section~\ref{sec:conclusion}. Throughout this paper, a flat $\Lambda$CDM cosmology was used, with $H_0 = 69$ km s$^{-1}$ Mpc$^{-1}$, $\Omega_M = 0.286$, and $\Omega_\Lambda = 0.714$ \citep{Bennett2014}.

%Using these datasets, we study the variability of these FSRQs over long and short timescales and their behavior at GeV and TeV energies during fast flares.

\section{Observations and Data Analysis}\label{sec:analysis}

\begin{deluxetable*}{lc|lcccc|ccc}
\tablecaption{Overview of the datasets presented in this paper.\label{tab:dataset}}
\tabletypesize{\scriptsize}
\tablecolumns{10}
\tablehead{
\multicolumn{2}{c|}{} \rule{0pt}{4ex} & \multicolumn{5}{c|}{\textit{Fermi}-LAT (HE gamma-ray)} & \multicolumn{3}{c}{VERITAS (VHE gamma-ray)}\\
\colhead{Source} & \multicolumn{1}{c|}{z}
& \colhead{Date Range} & \multirow{1}{1cm}{\centering Energy Range} & \multirow{1}{1cm}{\centering Time Binning} & \multirow{1}{0.85cm}{\centering No. Bins} & \multicolumn{1}{c|}{\multirow{1}{2.1cm}{\centering Flare Threshold (No. Flares)}}
    & \multirow{1}{1.15cm}{\centering Energy Threshold\tablenotemark{*}}    &   \colhead{Exposure}        & \multirow{1}{0.75cm}{\centering No. Obs.}\\
\multicolumn{2}{c|}{} & \multicolumn{5}{c|}{}\\
\multicolumn{2}{c|}{} & \colhead{[UT]} &  \colhead{[GeV]} & \colhead{[day]} & \colhead{} & \multicolumn{1}{c|}{[ph cm$^{-2}$ s$^{-1}$]} & \colhead{[GeV]} & \colhead{[hr]} & \colhead{}}
\startdata
3C 279          &   0.5362  &   2008-08-04 -- 2018-12-07   &   0.1-500 &   1   &   3471    & $4\times10^{-6}$ (10) &  200 &   54.4 &   139\\
PKS 1222+216    &   0.432\phn   &   2008-08-04 -- 2018-12-07   &   0.1-500 &   3  &   1158 & $5\times10^{-7}$  (11) & 110 &   34.7 &   \phn95\\
Ton 599         &   0.725\phn  &   2008-08-04 -- 2018-12-12   &   0.1-500 &   7  &   \phn512   & $5\times10^{-7}$ \phn(5) & 140 &   \phn8.8  &   \phn20\\
\enddata
\tablenotetext{*}{\textit{The energy threshold varies for different observations. A typical value is quoted for 3C~279 and the values during the VHE-detected flares are quoted for PKS~1222+216 and Ton~599.}}
\end{deluxetable*}

\subsection{VERITAS}\label{sec:VERITAS}

VERITAS is an array of four imaging atmospheric Cherenkov telescopes located in southern Arizona \citep[30$^\circ$ 40' N, 110$^\circ$ 57' W, 1.3 km above sea level;][]{Holder2011}.
%VERITAS is senstive to gamma rays from 85 GeV to greater than 30 TeV, with an angular resolution of 0.13 deg at 200 GeV and sensitivity equivalent to detecting a point source flux of 0.01 C.U. within 25 hours.
VERITAS preferentially performs observations of FSRQs when they exhibit an elevated flux in other wavebands, as a flare at TeV energies might also be occurring. The VERITAS observations of 3C~279, PKS~1222+216 and Ton~599 that were simultaneous with the HE flares considered here were taken in response to the elevated fluxes reported by \textit{Fermi}-LAT. VERITAS also carries out short monitoring observations of FSRQs. Because these sources are not believed to be strong emitters of TeV gamma rays except during flares, the primary aim of this monitoring is to self-trigger on serendipitous flares. For 3C~279 and PKS~1222+216, these observations provide VERITAS data corresponding to low states observed by \textit{Fermi}-LAT.

The total exposure of the VERITAS observations on each of the sources is reported in Table~\ref{tab:dataset}. The data were analyzed using a standard VERITAS data analysis package \citep{Maier2017} and cross-checked using an independent package \citep{Cogan08}. Boosted decision trees with soft selection cuts (appropriate for sources with a photon spectral index softer than $\Gamma \approx 3.5$) were used for separating gamma rays from background cosmic rays \citep{Krause17}. Preliminary analysis results of the VERITAS observations of 3C 279 and PKS 1222+216 in 2013 and 2014 were reported by \cite{2014HEAD...1410611E}. These are superseded by the more updated analyses reported here.

\subsection{\textit{Fermi}-LAT}\label{sec:fermi}

\textit{Fermi}-LAT detects gamma rays from 20 MeV to above 500 GeV using a pair-conversion technique \citep{Atwood2009}. \textit{Fermi}-LAT primarily operates in survey mode, during which it scans the entire sky every three hours.

We analyzed the Pass 8 data \citep{Atwood2013, Bruel2018} in the 10.3 year period starting on August 4, 2008 (MJD 54682.7), the start of the \textit{Fermi}-LAT all-sky survey, as reported in Table~\ref{tab:dataset}. We performed an unbinned likelihood analysis of the data using the LAT \texttt{Fermitools} 1.0.3 and instrument response functions {\texttt{P8R3\_SOURCE\_V2}}. The energy range from 0.1 GeV to 500 GeV was analyzed, and photons with zenith angle $> 90\degree$ were excluded to reduce contributions from the Earth's limb. For each source, the region of interest (ROI) considered was the circle of radius $10\degree$ surrounding the catalog source position. The background model consisted of, along with galactic ({\texttt{gll\_iem\_v06.fits}}) and isotropic ({\texttt{iso\_P8R3\_SOURCE\_V2.txt}}) diffuse emission models, all sources in the FL8Y catalog\footnote{\url{https://fermi.gsfc.nasa.gov/ssc/data/access/lat/fl8y/}} within a $20\degree$ circle surrounding the source. This is to ensure that the model would include gamma-ray emission from sources outside the ROI that could extend into the ROI due to the size of the point spread function of the LAT, especially at low energies. We excluded time ranges corresponding to solar flares and gamma-ray bursts in the ROI from the analysis.

%The galactic ({\texttt{gll\_iem\_v06.fits}}) and isotropic ({\texttt{iso\_P8R3\_SOURCE\_V2.txt}}) diffuse emission models were also included in the analysis.
% The fitting strategy consisted of letting the parameters of all significant sources, as provided by the FL8Y catalog, free to vary in the first iteration. Then sources with $TS<0$ were erased, and the sources found to be the least significant were systematically fixed up to a value of TS that increased with each additional iteration until convergence was reached.
When performing the likelihood fit, we iteratively fixed the parameters of the least significant sources (in increasing square powers of natural numbers up to TS equal to 25) until convergence was reached. Sources with TS values outside the allowed range, usually associated with flux parameter values close to zero, were removed from the model.
%The light curve and SED points in this manuscript were obtained having fixed the spectral parameters to the catalog values for the data provided in Figures \ref{fig:graylcs} and \ref{fig:global_seds}, and to the values found in dedicated analyses for the entire duration of each flare for data considered during these outbursts. Only the normalization parameters of the diffuse background models were left to vary.
When fitting individual light curve and SED points, the spectral parameters were kept fixed, either to their catalog values for global analyses or to values derived from an analysis of the full flare period for flare analyses, with the diffuse background model normalization parameters left free.
We checked that the background model we used is consistent with the 4FGL-DR2 catalog \citep{Abdollahi2020, Ballet2020} by comparing these two catalogs, finding no new bright, variable sources in the ROI of each of the three FSRQs that could significantly impact the analysis of our sources.

Since 3C~279 lies close to the ecliptic, the Sun and Moon contribute diffuse foreground emission in the ROI of this source during certain periods. This is at the level of $\sim 0.5\times 10^{-6}$ cm$^{-2}\,$s$^{-1}$ for the Sun \citep{Abdo2011} and $\sim 1\times 10^{-6}$ cm$^{-2}\,$s$^{-1}$ for the Moon \citep{Abdo2012}. The Sun's quiescent gamma-ray emission extends over a $20\degree$ radius, so this emission is partially degenerate with the diffuse backgrounds modeled in the likelihood fit. The Moon moves about $13\degree$ per day, so it appears within a time bin of a day as a strip. No contamination is expected during any of the flare states identified in Section~\ref{sec:flarealgorithm}, since both the Sun and Moon were more than $20\degree$ from 3C~279 during these periods.

% For 3C~279, there are multiple intervals for which the Sun is less than $20^{\circ}$ from the source and will, therefore, typically be contributing quiescent emission at the level of $\sim 0.5\times 10^{-6}$ cm$^{-2}\,$s$^{-1}$ \citep{Abdo2011}. None of these intervals occurred during the flare states identified in Section~\ref{sec:flarealgorithm}. Since the Sun's quiescent gamma-ray emission extends over a $20^{\circ}$ radius, this emission is partially accounted for by the diffuse background models in the likelihood fit.

To our knowledge, the maximum level of contamination due to the proximity of 3C~279 to the ecliptic has not previously been quantified. To do so, we generated extended templates for the Sun and the Moon for a 1-day time bin containing the closest approach of the Sun to 3C~279 during the period considered for our analysis \citep{2013ICRC...33.3106J}, that is, the bin containing October 9, 2018 when its annual occultation occurred \citep{Barbiellini2014}. 
During this time bin, the Sun reached a distance of $\approx 0.2^{\circ}$ from 3C~279. By coincidence, the Moon passed within $5\degree$ of the source during the same interval. The templates accounted for the expected extended emission from the Sun and Moon during a 1-day time bin. When we include these templates in the model file for the likelihood analysis for this bin we find that the flux of 3C~279 decreases by approximately $28\%$ with respect to that obtained when only the point sources in the ROI and the galactic and isotropic diffuse backgrounds are included. 
%This is the time bin for which the solar flux has the maximal impact on the flux from 3C~279.
The gamma-ray emission from the quiescent Sun and Moon is expected to vary with the solar cycle. In order to estimate the worst-case contamination, we also chose a selection of time bins during which only the Moon was present. When its diffuse template is included in the analysis, this results in a decrease of the 3C 279 flux by up to 49\% with respect to when the template is omitted.

For time bins in which the Sun or Moon is more than 5$^\circ$ from 3C~279, the flux of 3C~279, as returned by the likelihood analysis, does not change significantly when the solar and lunar templates are included. We conclude, therefore, that these contributions show no evidence of being statistically significant when deriving the spectral properties of 3C~279 for the time periods studied in this work.
%Any residual emission might not greatly affect the analysis of 3C~279, which is a very bright source, however, 
%We ask the reader to consider this caveat when making use of the data during these time periods. %(marked with a `$\ast$' in the data files provided). 
The Sun and Moon each come within $5\degree$ of 3C~279 for approximately 11-13 days per year. A more complete treatment, beyond the scope of this paper, could include the Sun and Moon as extended sources in the likelihood fits for these time bins to fully account for their emission.

\subsection{\textit{Swift}-XRT}\label{sec:xrt}

The X-Ray Telescope (XRT) on the Neil Gehrels {\it{Swift}} observatory is a grazing-incidence focusing X-ray telescope, and is sensitive to photons with energies between 0.2 and 10~keV \citep{Gehrels04, Burrows05}. \textit{Swift}-XRT observed PKS~1222+216 and Ton~599 during the VHE flares of those sources.

The \textit{Swift}-XRT data were extracted from the \textit{Swift} data archive and analyzed using \texttt{HEASoft v6.24}. The fluxes and flux errors were deabsorbed using the fixed total column density of Galactic hydrogen $N_H = 2.29 \times10^{20} \;\text{cm}^{-2}$ for PKS~1222+216 and $1.89 \times10^{20} \;\text{cm}^{-2}$ for Ton~599 \citep{Kalberla05,Willingale13} %\footnote{\url{https://www.swift.ac.uk/analysis/nhtot/index.php}}, $N_H$, 
and the photoelectric cross section $\sigma(E)$ to account for the effects of neutral hydrogen absorption. The deabsorbed X-ray spectrum was fitted with a broken power law model for PKS~1222+216 and a power law model for Ton~599.

% \begin{equation}
%     \label{deabs}
%     \frac{dN}{dE}_{deabs} = F e^{{N_H}{\sigma(E)}}.
% \end{equation}

% \begin{equation}\label{bknpowerlaw}
% \frac{dN}{dE} = \begin{cases} 
%       K E^{-\Gamma_1} & E \leq E_{break} \\
%       K (E_{break})^{\Gamma_2 - \Gamma_1} (\frac{E}{1\;\text{keV}})^{-\Gamma_1} & E > E_{break}
%   \end{cases}
% \end{equation}

% where the normalization $K$ and the photon indices $\Gamma_1$ and $\Gamma_2$ describe the spectrum at energies below and above the breaking energy $E_{break}$, respectively.

\subsection{\textit{Swift}-UVOT}\label{sec:UVOT}

The ultraviolet/optical telescope (UVOT) on the Neil Gehrels \textit{Swift} observatory is a photon counting telescope sensitive to photons with energies ranging from about 1.9 to 7.3 eV or 170 to 550 nm \citep{2005SSRv..120...95R}. %The UVOT spectral data were extracted from the \textit{Swift} data archive with the XRT data. 
\textit{Swift}-UVOT observed PKS~1222+216 and Ton~599 approximately concurrently with \textit{Swift}-XRT.

The UVOT data were extracted from the \textit{Swift} data archive and analyzed using \texttt{HEASOFT v6.28}. The counts from the sources and the background were extracted from regions of a radius of $5.0''$ centered on the position of the sources and nearby positions without any bright sources, respectively. The magnitude values of the sources were computed using \texttt{uvotsource}, and converted to fluxes using the zero-points given by \citet{Poole08}. Extinction corrections were applied following \citet{2009ApJ...690..163R}, using the reddening values $E(B-V)=$ 0.0199 and 0.0171 \citep{SandF2011} for PKS~1222+216 and Ton~599, respectively. 

\subsection{Steward Observatory}\label{sec:steward}

During the first decade of the \textit{Fermi} mission, the Steward Observatory of the University of Arizona obtained optical polarimetry, photometry, and spectra of the LAT-monitored blazars and \textit{Fermi} targets of opportunity (ToOs) using the SPOL CCD Imaging/Spectropolarimeter \citep{Smith2009}. We downloaded the spectrophotometric Johnson V and R band magnitudes from the Steward Observatory public archive\footnote{\url{http://james.as.arizona.edu/~psmith/Fermi/}}. These magnitudes were obtained by convolving the flux spectra between 4000 and 7550 {\AA} with a synthetic filter bandpass for the V or R band, summing the flux, and computing the magnitude difference with a comparison star. \citet{Smith2009} give the full details of the observations and data reduction. We then converted the magnitude for each bandpass to its equivalent energy flux. Six observations were taken of Ton~599 and two of PKS~1222+216 during their respective VHE flares. There was no significant variability during either event.

% \subsection{Owens Valley Radio Observatory (OVRO)}\label{sec:ovro}

% The 40 M Telescope at the Owens Valley Radio Observatory (OVRO) observes over 1800 blazars about twice per week as part of its \textit{Fermi} monitoring program \citep{Richards2011}. The reduced data are available online\footnote{\url{https://www.astro.caltech.edu/ovroblazars/index.php}}. OVRO observed Ton~599 once during its TeV flare, and does not monitor PKS~1222+216.

\section{\textit{Fermi}-LAT Flux Distributions}\label{sec:fluxdistributions}

\begin{figure*}[htbp]
\includegraphics[width=\textwidth]{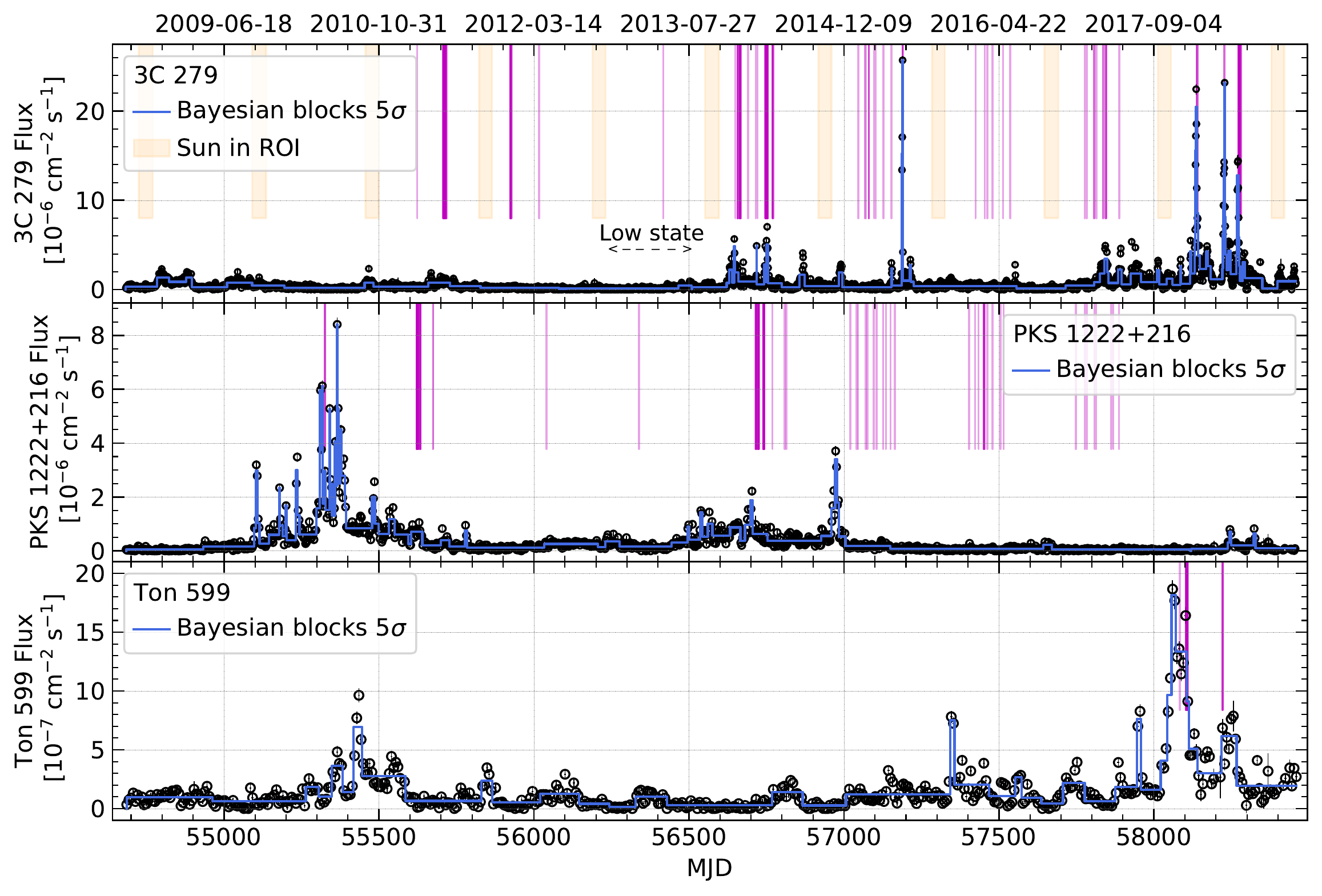}
\caption{\textit{Fermi}-LAT light curves of 3C~279 (top), PKS~1222+216 (middle), and Ton~599 (bottom). The flux points (black circles) are shown for 1, 3, and 7 day time bins for the three sources, respectively. $5\,\sigma$ Bayesian blocks are shown with blue lines. The time intervals in which VERITAS observed the sources are marked in magenta. For 3C~279, time intervals in which the Sun is less than $20^{\circ}$ from the source are shown in orange and a \textit{Fermi}-LAT low state from MJD 56230 -- 56465 (see Section~\ref{sec:spectra}) is marked with a dashed line.}
\label{fig:graylcs}
\end{figure*}

The \textit{Fermi}-LAT light curves of the three sources and the periods of the VERITAS observations are shown in Figure~\ref{fig:graylcs}. The LAT time binnings, reported in Table~\ref{tab:dataset}, were chosen for each source depending on its typical strength to avoid having an excessive number of bins with no detection.

\begin{figure*}
    \centering
    \includegraphics[width=\textwidth]{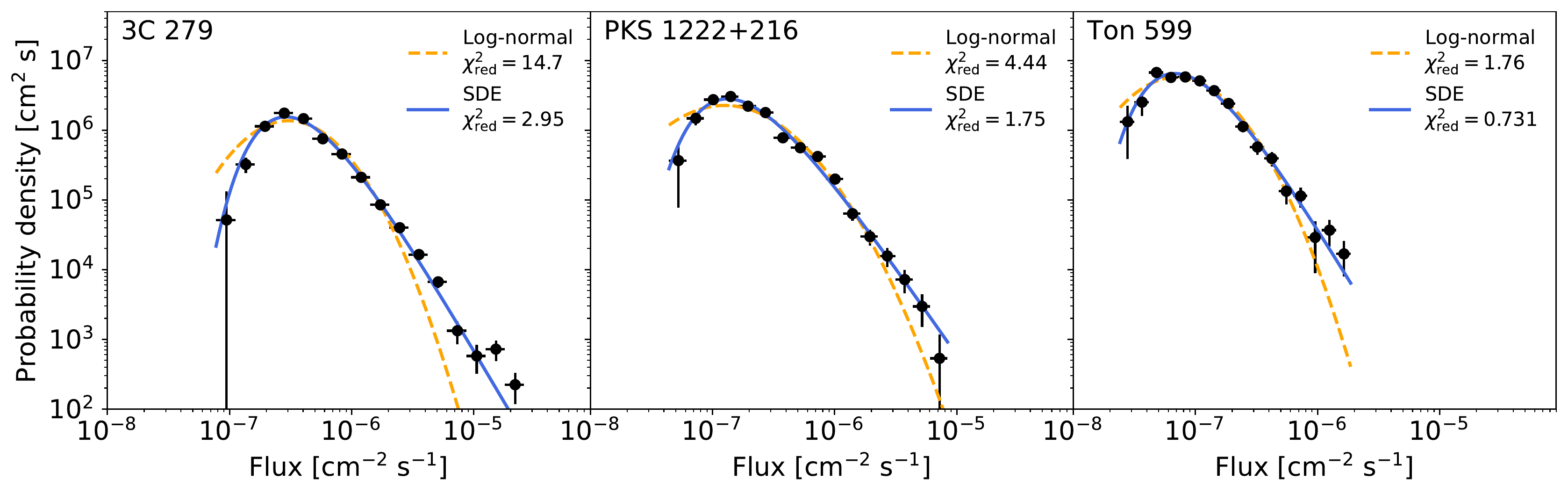}
    \caption{Flux distributions of the three FSRQs, scaled as probability densities. The distributions are fit with a log-normal PDF (dashed orange) and the stationary-state PDF corresponding to the SDE of \citet{Tavecchio2020} (solid blue). In all three cases, the SDE provides a better fit.}
    \label{fig:flux_distributions}
\end{figure*}

The distribution of the LAT fluxes observed from each of these FSRQs may provide a clue to the origin of the gamma-ray emission. The observed flux distributions of the three sources (scaled to form probability density histograms) are shown in Figure~\ref{fig:flux_distributions}. Time bins that have a test statistic (TS) less than 9 or that occur when the Sun is less than $20^{\circ}$ from the source were excluded.

To account for uncertainties from both the flux binning and the finite observation length, the flux histogram bin errors were calculated using a bootstrapping approach. 2,500 bootstrap samples were used, each consisting of the same number of flux points as the actual light curve. Each bootstrap sample was obtained by sampling from the set of actual flux points with replacement, so that a given flux point might be sampled multiple times or not at all.

In order to include the uncertainties of the individual flux points, an error term was added to each sampled point in each bootstrap sample, determined by randomly sampling from a Gaussian distribution with standard deviation equal to the measurement uncertainty of the respective sampled point. The bin errors were then defined as the standard deviations of the bin fluxes over all of the bootstrap samples binned using the same bins as the original dataset.

One form of flux distribution often used to describe blazars is the log-normal distribution \citep{Giebels09, Sinha17, Shah2018}. Log-normal distributions are of interest because they indicate the presence of an underlying multiplicative rather than additive physical process \citep{Aitchison1973}. Light curves with a log-normal flux distribution have an amplitude of variability linearly proportional to their mean flux.

On the other hand, \citet{Tavecchio2020} have proposed an alternative model based on an SDE with two terms modeling a deterministic tendency to return to equilibrium and stochastic fluctuations with amplitude proportional to the absolute flux level. The form of the SDE is motivated by an astrophysical scenario of stochastic disturbances perturbing a magnetically arrested accretion disk. In this model, the flux distribution is asymmetrical about a peak, falling off as a power law at high fluxes and exponentially at low fluxes, with the relative importance of the deterministic and stochastic components dictating the shape of the distribution. Figure~\ref{fig:flux_distributions} shows a comparison between the best-fit probability density functions (PDFs) corresponding to a log-normal distribution and the stationary state of the SDE proposed by \cite{Tavecchio2020}.
%These fluctuations have been estimated as the excess variance, $\sigma_{XS}$ \citep{Giebels09,Vaughan03}.

\begin{deluxetable*}{l|ccc|cccc}[htb]
\tablecaption{Best-fit parameters and goodness of fit ($\chi^2_{\rm{red}}$) for the log-normal and SDE PDF fits to the LAT flux distributions. The mean $\mu$ and standard deviation $\sigma$ of the exponentiation of the log-normal distribution and the equilibrium flux $\mu$ of the SDE distribution are normalized to $1\times10^{-7}$ ph cm$^{-2}$ s$^{-2}$. The peak flux of the SDE PDF is given by $X_\mathrm{max} = \mu \lambda / (\lambda + 2)$.
\label{tab:flux_distribution_fits}}
\tablehead{\colhead{Source} & \multicolumn{3}{c}{Log-normal} & 
\multicolumn{4}{c}{SDE}\\
 & \colhead{$\mu$} & \colhead{$\sigma$} & \colhead{$\chi^2_{\rm{red}}$} & \colhead{$\mu$} & \colhead{$\lambda$} & \colhead{$X_\mathrm{max}$} & \colhead{$\chi^2_{\rm{red}}$}}
\startdata
  3C 279   & 1.65$\pm$0.02   &  0.73$\pm$0.01   &  14.7 &  8.63$\pm$0.28 &  1.01$\pm$0.06 & 2.90$\pm$0.21 & 2.95\\
  PKS 1222+216    & 1.07$\pm$0.03   &  0.92$\pm$0.03   &  4.44  &  6.27$\pm$0.60 &  0.54$\pm$0.08 & 1.33$\pm$0.23 &  1.75\\
  Ton 599   & 0.21$\pm$0.04   &  0.75$\pm$0.03   &  1.76  &  2.11$\pm$0.19 &  0.94$\pm$0.15 & 0.68$\pm$0.13 & 0.73\\
\enddata
\end{deluxetable*}

The stationary-state PDF corresponding to the SDE \citep[Appendix A]{Tavecchio2020} is

\begin{equation}
    p(X) = \frac{(\lambda \mu)^{1 + \lambda}}{\Gamma(1 + \lambda)}\frac{e^{-\lambda \mu / X}}{X^{\lambda + 2}},
\end{equation}

\noindent where $X$ is a dimensionless random variable proportional to the flux, $\mu$ is a parameter representing the equilibrium value of $X$, $\lambda$ is a parameter representing the relative weight of the deterministic and stochastic terms, and $\Gamma$ is the gamma function. Here, $X$ was related to the flux by a proportionality constant of $1\times10^{-7}$ ph cm$^{-2}$ s$^{-1}$. The distribution peaks at $X_\mathrm{max} = \mu \lambda / (\lambda + 2)$. The stationary-state PDF is valid on timescales much longer than the timescale for the system to return to equilibrium, which is clearly the case for the ten-year periods considered here.

%The PDFs were fit to the histogram bins using the \texttt{scipy.optimize.curve\_fit} method\footnote{\url{https://docs.scipy.org/doc/scipy/reference/generated/scipy.optimize.curve_fit.html}} available in Python.
The PDFs were fit to the histogram bins using a nonlinear least-squares algorithm. The best-fit parameters and reduced $\chi^2$ values of the two models are reported in Table~\ref{tab:flux_distribution_fits}. In all three cases, the SDE PDF provides a better fit than the log-normal PDF. Both models have two free parameters. We verified that the preference for the SDE model is preserved if the histogram bins at the lowest fluxes, which might be affected by requiring light curve bins to have TS $>9$, are excluded from the fit.

The SDE model PDF is parameterized by the shape parameter $\lambda \equiv 2\theta/\sigma^2$, where $\theta$ and $\sigma$ are the coefficients of the deterministic and stochastic terms. These parameters can be interpreted by associating $1/\theta$ with the timescale of magnetic field accumulation in the accretion disk, while $\sigma$ is related to the dynamics of the perturbative processes. A large value of $\lambda$ therefore represents a high relative importance of the deterministic variability component compared to the stochastic one, while a small value indicates the opposite. To relate these timescales to the gravitational radii of the central supermassive black holes, $r_g = GM/c^2$, we adopt values of ${\sim}5\times10^8$, $6\times10^8$, and $3.5\times10^8 M_\odot$ for the black hole masses of 3C~279, PKS~1222+216, and Ton~599, respectively \citep{Hayashida2015, Farina2012, Liu2006}.

One can estimate $\sigma^2$ from the light curve using the expression \citep{Tavecchio2020}:

\begin{equation}
    \sigma^2 \simeq \frac{1}{n} \sum_{i=0}^{n} \frac{(X_i - X_{i - 1})^2}{X_{i - 1}^2 (t_i - t_{i - 1})},
\end{equation}

\noindent where $X_i$ is the scaled flux at time step $i$. Using this expression, we obtain $\sigma^2$ equal to 0.35, 0.16, and 0.062 day$^{-1}$, or 100, 200, and 800 $r_g / c$, for 3C~279, PKS~1222+216, and Ton~599, respectively. These values are consistent with the ${\gtrsim}100~r_g / c$ variability timescale injected into the jet by magneto-rotational instability in the accretion disk estimated in theoretical work \citep{Giannios2019}. Using the relation $1/\theta = 2/\lambda \sigma^2$, we can then constrain the physics of accretion flow in 3C~279, PKS~1222+216, and Ton~599 by estimating their magnetic flux accumulation timescales to be 200, 700, and 1800 $r_g / c$, respectively, within the magnetically-arrested disk scenario.

\section{Flare Selection}\label{sec:flarealgorithm}

Flare states were identified in the \textit{Fermi}-LAT data using the following procedure:

\begin{enumerate}
    \item Segment the data using Bayesian blocks. We set the false positive rate $p_0$ to the value equivalent to $5\,\sigma$ using Equation 13 of \citet{Scargle13}.
    \item Choose a flux threshold above which the blocks are designated as flaring.
    \item Designate each contiguous set of flare blocks as an individual flare state and all non-flare blocks as the quiescent state. 
\end{enumerate}

This empirical procedure reflects a picture of individual flares superimposed on a constant quiescent background, but identifies them purely as states of elevated flux, making no explicit assumptions about the flares' shape or spectra. Due to its basis on Bayesian blocks, it guarantees that states identified as flares have flux significantly greater than the states surrounding them.

The flux threshold to identify flares must be tuned on a source-by-source basis. Choosing the flux threshold to identify flares involves a trade-off between ensuring that dimmer flares are selected and avoiding misidentifying fluctuations in the quiescent background as flares. In addition, because the sources differ in average flux, the threshold must necessarily vary on an absolute level from source to source. Performing the flare selection procedure with the flare selection thresholds listed in Table~\ref{tab:dataset} results in 10 flares selected for 3C 279, 11 for PKS 1222+216, and 5 for Ton 599, listed in Table~\ref{tab:flare_states}. We set the threshold low enough for each source to ensure that all flares that triggered VERITAS observations were selected.

\begin{deluxetable}{crrrc}[ht]
\tablecaption{\textit{Fermi}-LAT flares selected using the algorithm given in Section~\ref{sec:flarealgorithm}. For each enumerated flare, the date range in MJD, approximate calendar date, number of Bayesian blocks, and amount of VHE gamma-ray exposure taken by VERITAS (if any) are provided. All of the times in the date ranges given in Table~\ref{tab:dataset} but not listed here are considered to be quiescent.
\label{tab:flare_states}}
\tablehead{\colhead{\#} & \colhead{Date Range (MJD)} & \colhead{Approx. Date} & \colhead{Blocks} & \colhead{VHE Exp.}}
\startdata
\multicolumn{5}{c}{3C 279}\\
\hline
1 & 56645.66 - 56647.66 & Dec 2013 & 1 & -\\
2 & 56717.66 - 56718.66 & Mar 2014 & 1 & -\\
3 & 56749.66 - 56754.66 & Apr 2014 & 1 & 6.79 hr\\
4 & 57186.66 - 57190.66 & Jun 2015 & 3 & 1.00 hr\\
5 & 58116.66 - 58119.66 & Dec 2017 & 1 & -\\
6 & 58130.66 - 58141.66 & Jan 2018 & 4 & 1.38 hr\\
7 & 58168.66 - 58173.66 & Feb 2018 & 1 & -\\
8 & 58222.66 - 58230.66 & Apr 2018 & 5 & 0.83 hr\\
9 & 58239.66 - 58247.66 & May 2018 & 1 & -\\
10 & 58268.66 - 58275.66 & Jun 2018 & 2 & 3.95 hr\\
\hline
\multicolumn{5}{c}{PKS 1222+216}\\
\hline
1 & 55096.66 - 55114.66 & Sep-Oct 2009 & 3 & -\\
2 & 55144.66 - 55201.66 & Nov-Dec 2009 & 5 & -\\
3 & 55231.66 - 55594.66 & 2010 & 27 & -\\
4 & 55603.66 - 55639.66 & Feb-Mar 2011 & 4 & 5.38 hr\\
5 & 55777.66 - 55783.66 & Aug 2011 & 1 & -\\
6 & 56494.66 - 56500.66 & Jul 2013 & 1 & -\\
7 & 56536.66 - 56665.66 & Sep 2013 & 5 & -\\
8 & 56680.66 - 56752.66 & Jan-Apr 2014 & 3 & 15.53 hr\\
9 & 56926.66 - 57004.66 & Sep-Dec 2014 & 5 & -\\
10 & 58243.66 - 58249.66 & May 2018 & 1 & -\\
11 & 58321.66 - 58327.66 & Jul 2018 & 1 & -\\
\hline
\multicolumn{5}{c}{Ton 599}\\
\hline
1 & 55417.66 - 55445.66 & Aug-Sep 2010 & 1 & -\\
2 & 57342.66 - 57356.66 & Nov 2015 & 1 & -\\
3 & 57944.66 - 57958.66 & Jul 2017 & 1 & -\\
4 & 58042.66 - 58140.66 & Oct 2017 - Jan 2018 & 5 & 8.30 hr\\
5 & 58217.66 - 58266.66 & Apr-May 2018 & 1 & 2.00 hr\\
\enddata
\end{deluxetable}

Because the flux distributions are best fit by the single-component SDE model PDF, it is not natural to calculate a duty cycle of flares based on a division into baseline and flaring components \citep[e.g.][]{Resconi2009}. The amount of time spent in the highest-flux states can be estimated directly from the flux distribution by defining the ``typical" flux as the peak of the PDF, given in Table~\ref{tab:flux_distribution_fits}. 3C~279, PKS~1222+216, and Ton~599 have flux greater than 5 (10) times the typical flux 12\% (4\%), 19\% (8\%), and 13\% (4\%) of the time, respectively.

Our flare selection flux thresholds for 3C~279 and Ton~599 are comparable at 13.8 and 11.8 times their typical fluxes, consistent with their similar values of the PDF shape parameter $\lambda \approx 1$. For PKS~1222+216, our threshold is 3.8 times the typical flux. This source has a lower value of $\lambda \approx 0.5$, with a correspondingly harder power law of the flux distribution at high fluxes. This is perhaps reflected in the long epochs of high flux seen in this source's light curve, such as its Flare~3 in 2010 which is approximately a year in duration (Table~\ref{tab:flare_states}). A relatively low threshold was therefore needed to also capture the smaller flares of the approximately weekly timescales that typically trigger VERITAS observations, consistent with the other two sources.

\section{Daily and sub-daily variability}\label{sec:flare_profiles}

In order to deduce the smallest variability time around the rising and decaying periods of each flare selected according to the algorithm described in Section \ref{sec:flarealgorithm}, we extracted sub-daily light curves of the three sources in time bins ranging from 12 hours down to 1.5 hours for the brightest source, 3C~279. Starting from daily time bins, we refined the light curve iteratively by splitting the time bin duration until each bin had a TS of $\gtrsim50$ or until further refinement would not change the local trend of the light curve. For PKS~1222+216 and Ton~599, the minimum bin sizes were 12 and 6 hours, respectively.
%down to the smallest significant time-bin sizes.
% Moved to Fermi-LAT analysis section
%The spectral parameters in all time bins for each flare were fixed to the parameters derived from an analysis of the full flare period.
%The bins ranged from 12 hours down to 1.5 hours for the brightest source, 3C~279. For PKS~1222+216 and Ton~599, the minimum bin sizes were 12 and 6 hours, respectively.

To characterize the flares with multiple peaks we used a sum of exponential profiles \citep{Valtaoja1999, Abdo2010}, $F_i$, where each one has the form:

\begin{equation} \label{eq:flareprofile}
F_i(t) = \begin{cases} 
      F_{0_i}\, e^{(t-t_{\rm{peak}_i})/t_{\rm{rise}_i}}, & t\leq t_{\rm{peak}_i}\\
      F_{0_i}\, e^{-(t-t_{\rm{peak}_i})/t_{\rm{decay}_i}}, & t> t_{\rm{peak}_i}.
   \end{cases}
\end{equation}

For flares with a single peak we used:

\begin{equation} \label{eq:flareprofileconst}
F(t) = \begin{cases} 
      F_0\, e^{(t-t_{\rm{peak}})/t_{\rm{rise}}}+F_\mathrm{const}, & t\leq t_{\rm{peak}}\\
      F_0\, e^{-(t-t_{\rm{peak}})/t_{\rm{decay}}}+F_\mathrm{const}, & t> t_{\rm{peak}},
   \end{cases}
\end{equation}

\noindent including a constant term to avoid having a bias towards large rise and decay timescales, which is minimal when multiple peaks are included.

The fitting procedure started by considering a single peak characterized by Equation (\ref{eq:flareprofileconst}). In order to evaluate the possibility of adding a second peak, a fit to the sum of two exponential profiles, as given by Equation (\ref{eq:flareprofile}), was performed and compared against the one-peak scenario using the reduced $\chi^2$ method. The two-peak model was then taken when an improvement was observed over the one-peak function. More peaks were then added following a similar procedure until a reasonable reduced $\chi^2$ value was reached, or when the best fit values obtained no longer provided relevant information for constraining the variability timescales of the sources under study. At each step, human judgment was used to initialize the profile positions and determine by eye that the fits made sense. The peaks were not required to match the Bayesian blocks used for flare selection, which were defined using the coarsely binned light curves.

\begin{figure*}[htbp]
\centering
\includegraphics[width=.95\textwidth,keepaspectratio]{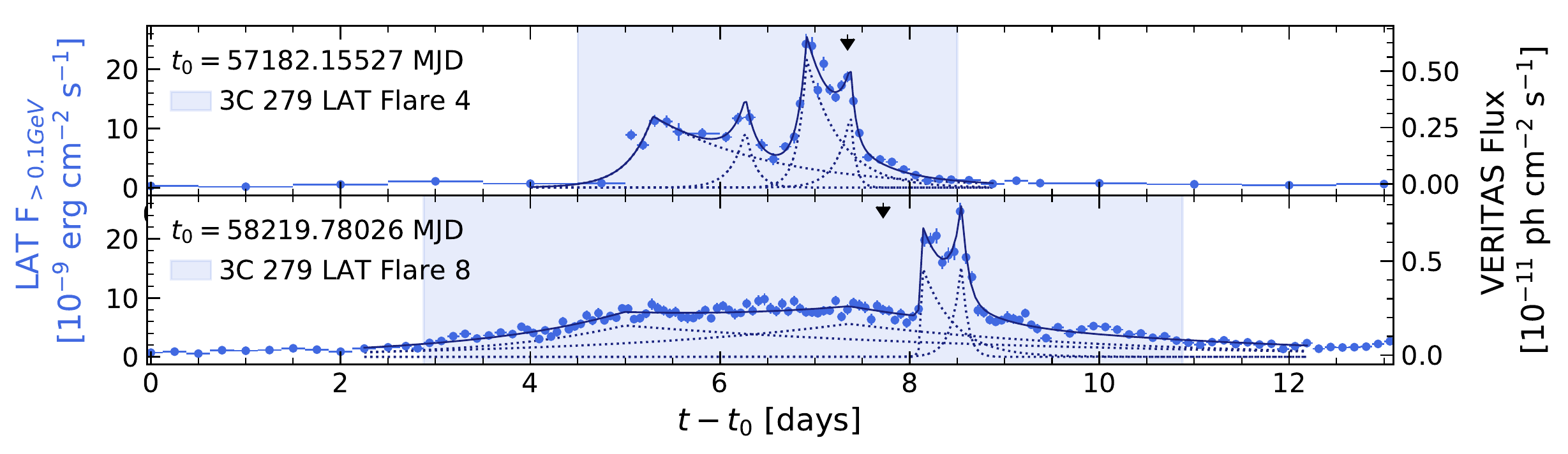}
\includegraphics[width=.95\linewidth,keepaspectratio]{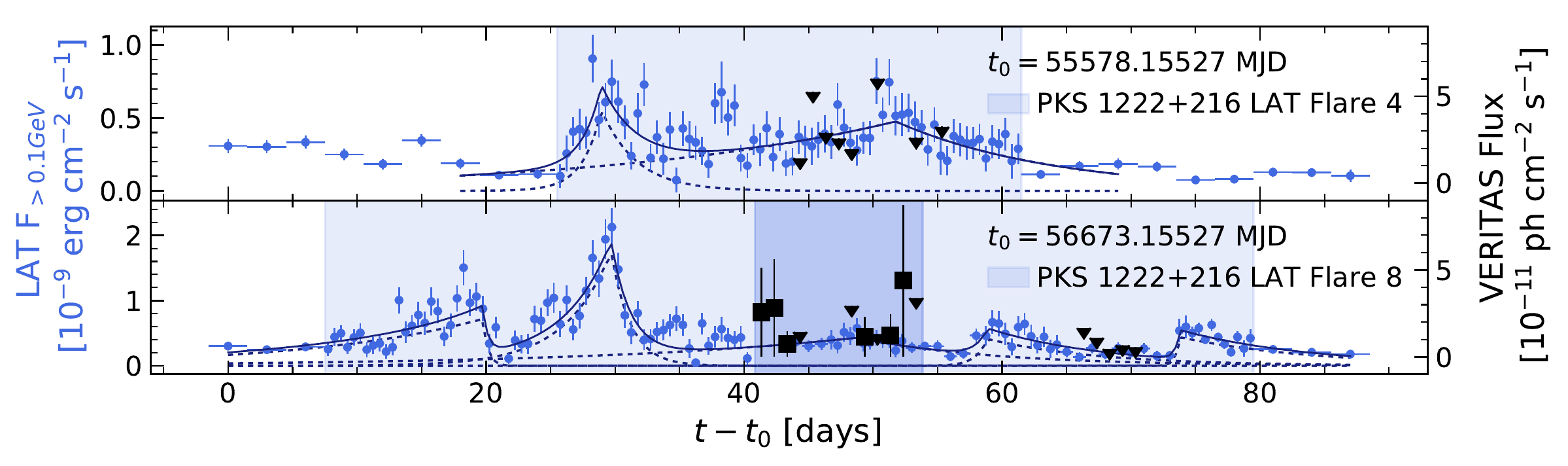}
\includegraphics[width=.95\linewidth,keepaspectratio]{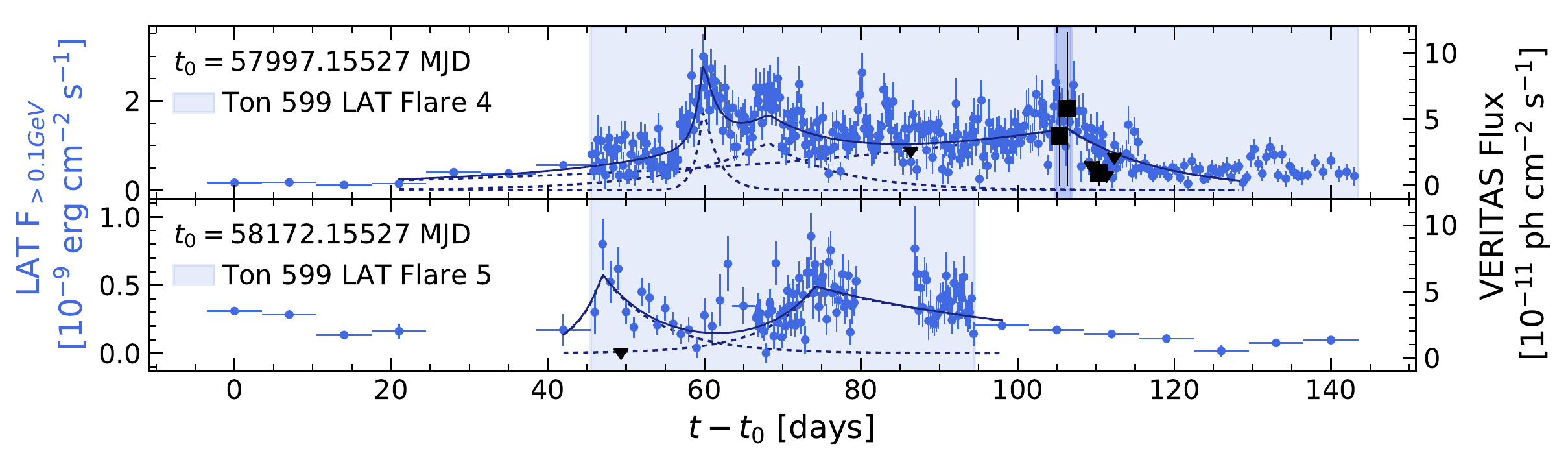}
\caption{LAT daily and sub-daily light curves (blue points) around selected flaring episodes (light shaded areas). The dotted blue lines show the fitted exponential profiles, with their sums shown in solid blue. The dark shaded areas indicate the periods considered for the SED modeling (Section~\ref{sec:sed_modeling}). The VERITAS data points and 95\% upper limits are shown as black squares and downwards arrows.}\label{fig:flareprofiles}
\end{figure*}

The flare profiles of the three sources are shown in Figure~\ref{fig:flareprofiles}. Two selected flares of 3C~279 are shown, as are the two flares each of PKS~1222+216 and Ton~599 that were observed by VERITAS. Profiles of all ten flares of 3C~279 are provided in Appendix~\ref{appendix:3c279flareprofiles}. In order to illustrate when VERITAS observed the source relative to the LAT flare peaks, the VERITAS daily-binned light curves for each of the flares are also shown in Figure~\ref{fig:flareprofiles}.

The fit results for the three sources are reported in Tables~\ref{tab:flareprofiles_3c279}, \ref{tab:flareprofiles_pkston}, and \ref{tab:flareprofile_3C279}. For 3C~279, each flare lasted between one and eleven days and consisted of between one and four separately resolved components, modeled using exponential profiles. Twenty-four distinct components are resolved within the ten flares. The rise and decay times range from timescales of days to less than one hour. The smallest resolved variability timescale was $36\pm13$ minutes, which occurred around MJD 58227.945, during the rising period of Flare~8 (MJD 58222.655 --  58230.655), indicated in boldface in Table~\ref{tab:flareprofiles_3c279} and Table~\ref{tab:flareprofile_3C279}.

\begin{deluxetable*}{ccccc}
\tablecaption{Results of the LAT flare profile fits for 3C~279, with flare timescales in \textit{minutes}.\label{tab:flareprofiles_3c279}}
%\tabletypesize{\scriptsize}
\tablehead{
\colhead{} & \colhead{Amplitude ($F_0$)} & \colhead{$t_{\rm{peak}}$} &  \colhead{$t_{\rm{rise}}$} & \colhead{$t_{\rm{decay}}$}\\
\colhead{} & \colhead{($10^{-9} $ erg cm$^{-2}\,$s$^{-1}$)}    & \colhead{(MJD)}  &  \colhead{(min)}  & \colhead{(min)}%\tablenotemark{a}}
}
\startdata
\multicolumn{5}{c}{Flare 4 (MJD 57186.655 -- 57190.655): $\chi^2/$d.o.f.= 77.31/19 = 4.07} \\\hline
& 12.07 $\pm$ 0.67 & 57187.446 $\pm$ 0.031 & 378 $\pm$ 46 & 1784 $\pm$ 147\\
& \phn9.79 $\pm$ 2.29 & 57188.425 $\pm$ 0.028 & 216 $\pm$ 101 & 155 $\pm$ 64\\
& 21.72 $\pm$ 1.59 & 57189.069 $\pm$ 0.008 & 137 $\pm$ 18 & 512 $\pm$ 55\\
& 12.41 $\pm$ 1.30 & 57189.532 $\pm$ 0.010 & 220 $\pm$ 63 & 77 $\pm$ 25\\\hline
\multicolumn{5}{c}{Flare 8 (MJD 58222.655 -- 58230.655): $\chi^2/$d.o.f.= 177.25/106 = 1.67} \\\hline
& \phn5.29 $\pm$ 1.29 & 58224.773 $\pm$ 0.105 & 1996 $\pm$ 716 & 5899 $\pm$ 4035\\
& 17.70 $\pm$ 2.01 & 58227.945 $\pm$ 0.004 & \bf{36 $\mathbf{\pm}$ 13} & 329 $\pm$ 131\\
& 16.42 $\pm$ 1.87 & 58228.323 $\pm$ 0.012 & 140 $\pm$ 54 & 115 $\pm$ 48\\
& \phn5.59 $\pm$ 1.69 & 58227.139 $\pm$ 0.133 & 3816 $\pm$ 1450 & 4077 $\pm$ 2080\\\hline
\enddata
\tablecomments{The smallest variability time found is indicated in boldface.}
\end{deluxetable*}

\begin{deluxetable*}{ccccc}
\tablecaption{Results of the LAT flare profile fits for PKS~1222+216 and Ton~599, with flare timescales in \textit{days}.\label{tab:flareprofiles_pkston}}
%\tabletypesize{\scriptsize}
\tablehead{
\colhead{} & \colhead{Amplitude ($F_0$)} & \colhead{$t_{\rm{peak}}$} &  \colhead{$t_{\rm{rise}}$} & \colhead{$t_{\rm{decay}}$}\\
\colhead{} & \colhead{($10^{-9} $ erg cm$^{-2}\,$s$^{-1}$)}    & \colhead{(MJD)}  &  \colhead{(days)}  & \colhead{(days)}%\tablenotemark{a}}
}
\startdata
\multicolumn{5}{c}{PKS~1222+216}\\\hline
\multicolumn{5}{c}{Flare 4 (MJD 55603.7 -- 55639.7): $\chi^2/$d.o.f.= 102.25/69 = 1.48} \\\hline
& 0.56 $\pm$ 0.09 & 55607.1 $\pm$ 0.3 & 1.5 $\pm$ 0.5 & 2.7 $\pm$ 0.8\\
& 0.48 $\pm$ 0.03 & 55629.9 $\pm$ 1.2 & 22.3 $\pm$ 3.5 & 12.1 $\pm$ 2.2\\\hline
\multicolumn{5}{c}{Flare 8 (MJD 56680.7 -- 56752.7): $\chi^2/$d.o.f.= 166.40/104 = 1.60} \\\hline
& 0.72 $\pm$ 0.09 & 56692.9 $\pm$ 0.1 & 13.6 $\pm$ 2.7 & 0.4 $\pm$ 0.3\\
& 1.75 $\pm$ 0.17 & 56702.8 $\pm$ 0.2 & 3.6 $\pm$ 0.6 & 1.5 $\pm$ 0.3\\
* & 0.43 $\pm$ 0.05 & 56721.9 $\pm$ 1.6 & 19.9 $\pm$ 8.5 & \textbf{10.4 $\pm$ 6.2}\\
& 0.41 $\pm$ 0.10 & 56732.0 $\pm$ 0.4 & 0.9 $\pm$ 0.8 & 8.6 $\pm$ 3.6\\
& 0.44 $\pm$ 0.06 & 56746.9 $\pm$ 0.1 & 0.3 $\pm$ 0.3 & 10.3 $\pm$ 2.5\\\hline
\multicolumn{5}{c}{Ton~599}\\\hline
\multicolumn{5}{c}{Flare 4 (MJD 58042.7 -- 58140.7): $\chi^2/$d.o.f.= 456.78/296 = 1.54} \\\hline
& 1.89 $\pm$ 0.29 & 58057.1 $\pm$ 0.2 & 1.1 $\pm$ 0.3 & 1.9 $\pm$ 0.6\\
& 1.06 $\pm$ 0.11 & 58065.4 $\pm$ 1.0 & 11.9 $\pm$ 2.6 & 9.0 $\pm$ 2.1\\
* & 1.37 $\pm$ 0.06 & 58103.5 $\pm$ 0.8 & 47.0 $\pm$ 4.7 & {\bf 11.8} $\pm$ {\bf 1.1}\\\hline
\multicolumn{5}{c}{Flare 5 (MJD 58217.7 -- 58266.7): $\chi^2/$d.o.f.= 153.01/96 = 1.59} \\\hline
& 0.57 $\pm$ 0.12 & 58219.2 $\pm$ 1.3 & 3.5 $\pm$ 2.8 & 7.3 $\pm$ 2.1\\
& 0.48 $\pm$ 0.03 & 58246.3 $\pm$ 0.7 & 6.6 $\pm$ 1.7 & 34.9 $\pm$ 5.6\\
\enddata
\tablecomments{The flare components coincident with VHE flares are marked with a *, with corresponding smallest variability times indicated in boldface.}
\end{deluxetable*}

For PKS~1222+216 and Ton~599, the variability timescales were of the order of days. Notably, for both sources, the fastest variability did not occur during the detected VHE flares. The shortest variability timescale observed by LAT during the VHE flare of PKS~1222+216 was $10.4 \pm 6.2$ days, which was the decay timescale of the coincident flare component. The shortest variability timescale of Ton~599 observed by LAT during its VHE flare was $11.8 \pm 1.1$ days, which also was the coincident flare component's decay timescale. In the case of Ton~599, the VERITAS detection occurred over a period of 2 days, after which the observed VHE flux became insignificant. No significant intra-flare variability was observed by \textit{Fermi}-LAT or VERITAS during either event. Therefore, in the remainder of this work, we take the most constraining variability timescales during the VHE flares of PKS~1222+216 and Ton~599 to be 10 and 2 days, respectively.

\begin{figure*}
    \includegraphics[width=\textwidth]{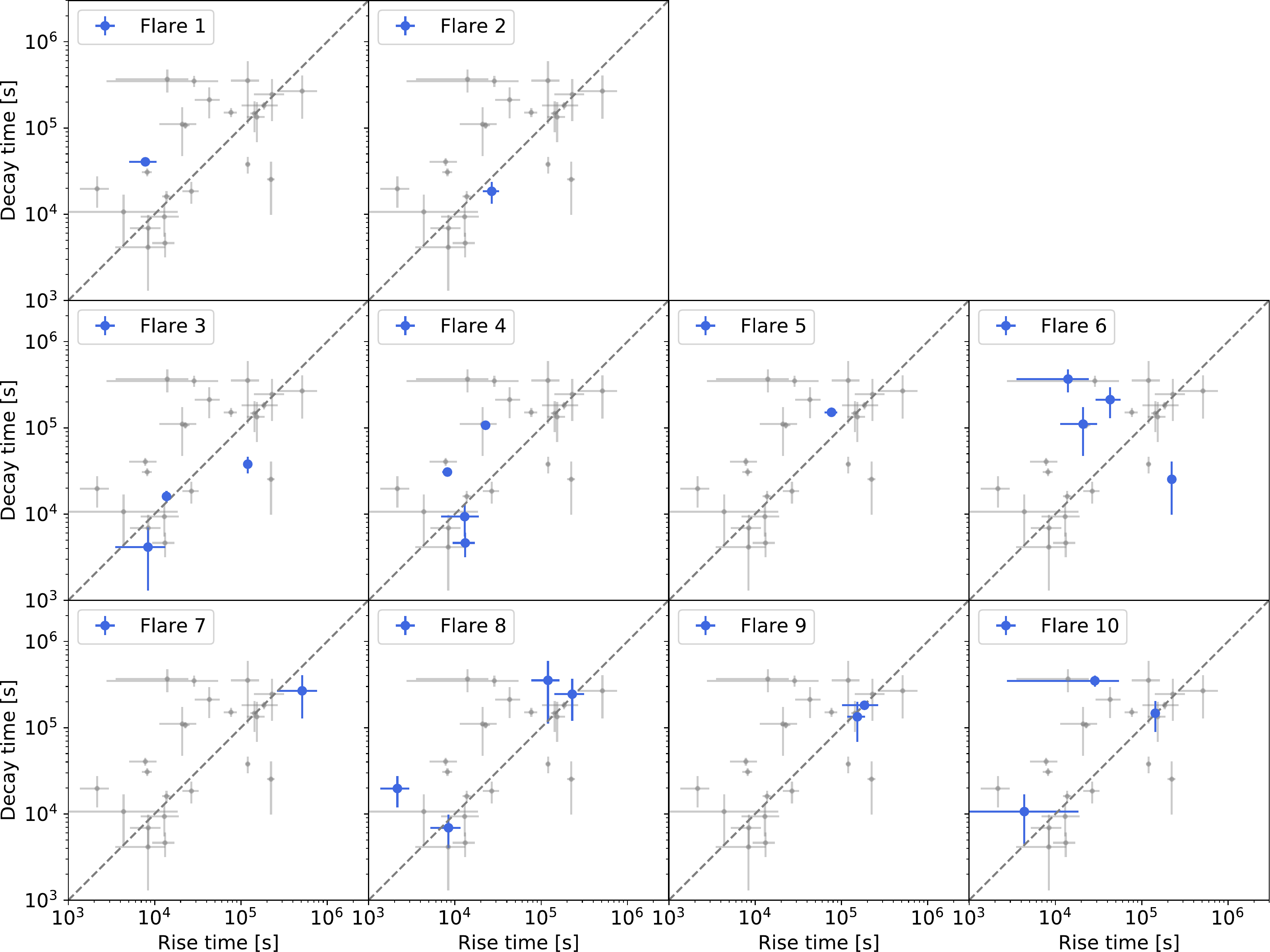}
    \caption{Decay time vs. rise time for each of the resolved exponential components in the flares of 3C~279. The points corresponding to all of the components are shown in gray. The dashed reference line shows where the rise and decay times are equal.}
    \label{fig:flare_profile_timescales}
\end{figure*}

% \begin{figure*}[htbp]
% \centering
% \includegraphics[width=0.48\textwidth]{Ton599_lat_flareprofile_2010.jpg}
% \includegraphics[width=0.50\textwidth]{Ton599_lat_flareprofile_2017.jpg}
% \caption{Daily light curves of Ton~599 during two flaring states in 2010 and 2017. The solid yellow lines show the fitted exponential profiles (Eq.~\ref{eq:flareprofile}).}\label{fig:flareprofileton599}
% \end{figure*}

%Investigating the properties of individual bright flares gives insight into the mechanisms causing the gamma-ray emission. The sub-daily lightcurves of the ten brightest flares of 3C~279 are shown in Figure~\ref{fig:flareprofile3C279}. 

The symmetry or asymmetry of flares can provide information on the timescales of the particle acceleration and cooling processes in the emission region \citep[e.g.][]{Abdo2010}. If the cooling time is longer than the light travel time through the emission region, the decay time will be longer than the rise time, producing an asymmetric flare. If the cooling time is shorter than the light travel time, the flare will appear more symmetrical. Flares with a slow rise and fast decay may be produced by relativistic magnetic reconnection \citep{Petropoulou2016}.

\begin{figure}
    \includegraphics[width=0.45\textwidth]{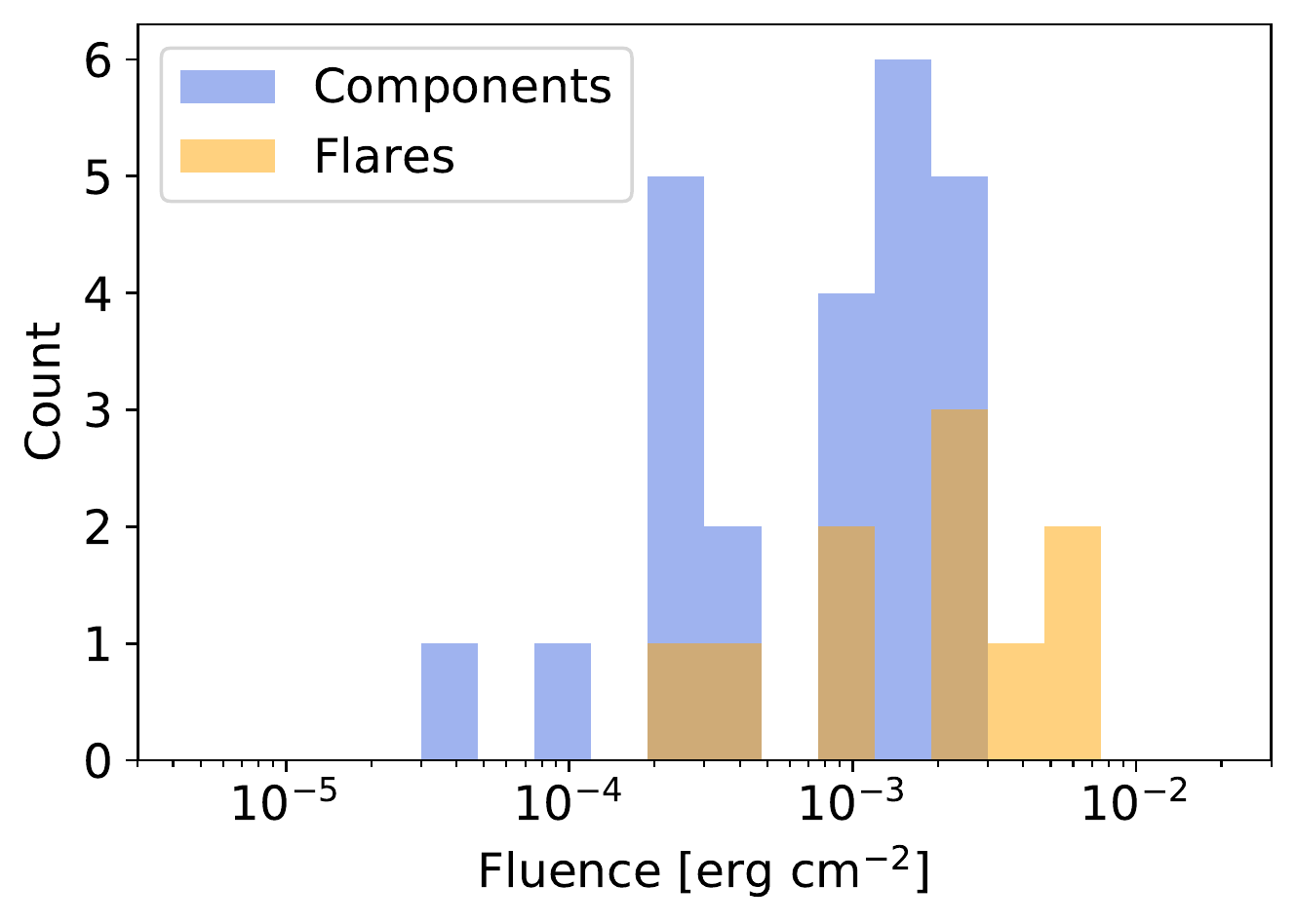}
    \caption{Fluence distributions of the twenty-four resolved flare components and ten flares of 3C~279.}
    \label{fig:flare_fluences}
\end{figure}

Figure~\ref{fig:flare_profile_timescales} shows the fitted rise and decay times for each of the exponential flare components of 3C~279. No clear trend in the flare asymmetry is observable, whether overall, among components within a single flare, or between the components belonging to different flares. Both longer decay times and longer rise times are observed, and many flares appear symmetric. A Wilcoxon signed-rank test \citep{Wilcoxon1945} finds no significant preference ($p=0.178$) for flares to have a faster rise time than decay time rather than the reverse. These findings are consistent with previous studies of gamma-ray flares in bright \textit{Fermi} blazars \citep[e.g.][]{Abdo2010, Roy2019}.

Models of blazar flares powered by relativistic reconnection predict that flare components produced by large, non-relativistic plasmoids should have similar fluences to components produced by small, relativistic ones, so that flare components should have similar fluence regardless of their variability timescales \citep{Petropoulou2016}. Figure~\ref{fig:flare_fluences} shows the distributions of fluences of the components of the ten flares and the twenty-four individual flare components of 3C~279. The fluence $\mathcal{F}$ of a flare with exponential components $F_i$ is given by:

\begin{equation}
    \mathcal{F} = \sum_i F_{0_i} (t_\mathrm{rise} + t_\mathrm{decay}).
\end{equation}

For 3C~279 Flares 1, 2, 5, and 7, the best fit is given by a single component plus a constant baseline flux. In these cases, the baseline flux is included in the fluence estimate for consistency with the other flares, approximating the flare duration as $t_\mathrm{rise} + t_\mathrm{decay}$, so that the fluence is given by:

\begin{equation}
    \mathcal{F} = (F_0 + F_\mathrm{const}) (t_\mathrm{rise} + t_\mathrm{decay}).
\end{equation}

The median flare fluence is $2.1\times10^{-3}$ erg cm$^{-2}$ and the median component fluence is $0.85\times10^{-3}$ erg cm$^{-2}$. The observed component fluences range over about one order of magnitude, as do the flare amplitudes, while the rise and decay timescales span about two orders of magnitude. These dynamic ranges are generally compatible with the expectations for plasmoid-powered flares derived from particle-in-cell simulations of relativistic magnetic reconnection \citep{Petropoulou2016}.

\begin{figure*}[htbp]\centering
\includegraphics[width=0.4\textwidth]{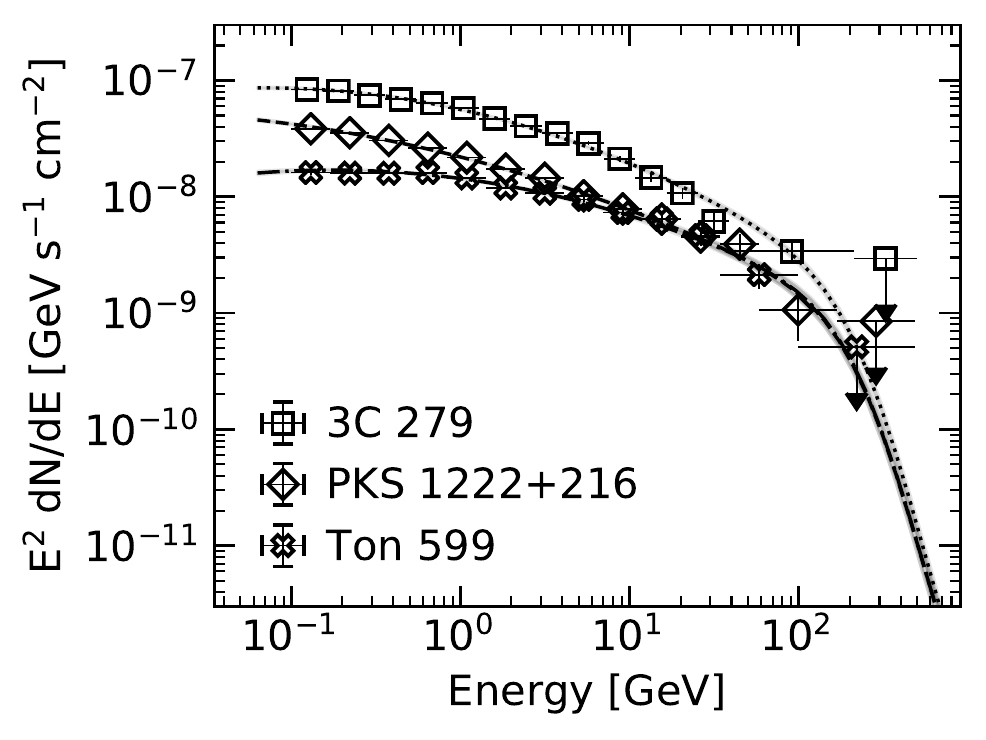}\hspace{1cm}
\includegraphics[width=0.4\textwidth]{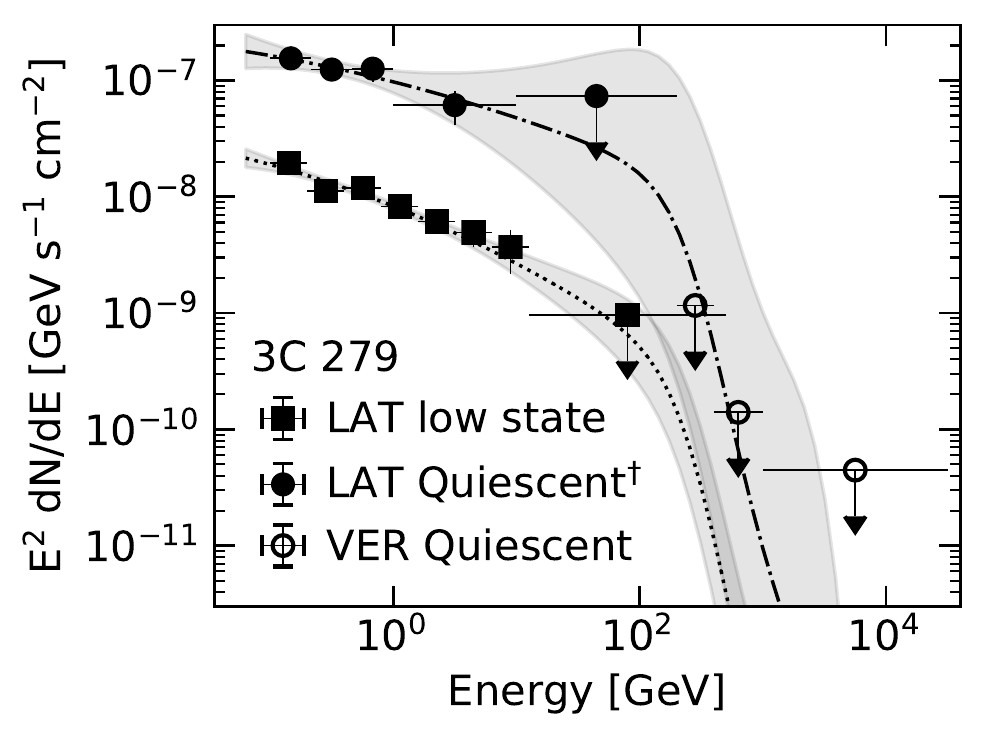}
\caption{{\sl Left:} Global \textit{Fermi}-LAT spectra for 3C~279, PKS~1222+216, and Ton~599. The LAT spectra are extrapolated to the VERITAS energy range, incorporating EBL absorption. {\sl Right:} Two baseline states of 3C~279. The \textit{Fermi}-LAT spectrum corresponding to the multiwavelength low state (MJD 56230--56465; see Figure~\ref{fig:graylcs}), is shown by filled squares. The strictly simultaneous \textit{Fermi}-LAT spectrum and VERITAS spectral upper limits during the quiescent state are shown by black filled and unfilled circles and contours. The symbol ``$^{\dagger}$" indicates that the LAT spectrum corresponds to data strictly simultaneous with VERITAS observations. Downward arrows show 95\% confidence level upper limits.}
\label{fig:global_seds}
\end{figure*}

The long-term gamma-ray variability study of the three FSRQs presented here is compatible with the extensive flare characteristics study carried out recently by \cite{Meyer2019} on the brightest flares detected by \textit{Fermi}-LAT. A similar Bayesian blocks analysis was carried out to identify flares and look for variability on sub-hour timescales. Consistent with their findings, we find sub-hour-scale variability in 3C~279, where it was possible to resolve flares in finer time bins, suggesting that extremely compact emission regions may be present within the jet.

% \begin{deluxetable*}{ccccc}
% \tablecaption{Results of the LAT flare profile fits for Ton 599.\label{tab:flareprofileton599}}
% %\tabletypesize{\scriptsize}
% \tablehead{
% \colhead{Amplitude ($F_0$)} & \colhead{$t_{\rm{peak}}$} &  \colhead{$t_{\rm{rise}}$} & \colhead{$t_{\rm{decay}}$}  & \colhead{Constant ($F_{\rm{const.}}$)}\\
% \colhead{($\times10^{-7} $ cm$^{-2}\,$s$^{-1}$)}    & \colhead{(MJD)}  &  \colhead{(days)}  & \colhead{(days)} & \colhead{($\times10^{-7} $ cm$^{-2}\,$s$^{-1}$)}%\tablenotemark{a}}
% }

% \startdata
% \multicolumn{5}{c}{Flare 1 (MJD 54952.65530 -- 55822.65530): $\chi^2/$d.o.f.= xx/xx = x.xx} \\\hline
% 3.93 $\pm$ xx & 55421 $\pm$ xx & {\bf 0.28} $\pm$ xx & 31.03 $\pm$ xx & xx $\pm$ xx \\\hline
% \multicolumn{5}{c}{Flare 2 (MJD 57772.65530 -- 58657.65530): $\chi^2/$d.o.f.= xx/xx = x.xx}  \\\hline
% 9.13 $\pm$ xx & 57949 $\pm$ xx & 2.06 $\pm$ xx & 6.23 $\pm$ xx & xx $\pm$ xx \\\hline
% \enddata
% \tablecomments{The smallest variability times found are indicated in boldface.}
% \end{deluxetable*}

\section{Gamma-ray Spectra} \label{sec:spectra}

Figure~\ref{fig:global_seds} shows the LAT energy spectra corresponding to the entire data sets of each of the three sources, along with the VERITAS spectral upper limits for 3C~279. The best-fit spectral parameters are reported in Appendix~\ref{appendix:spectral_fit_parameters}. Since all three sources were best fit by a log-parabola model in the 4FGL catalog \citep{Abdollahi2020}, we fit the LAT spectra with this model, parametrized as 

\begin{equation}
    \frac{dN}{dE} = N_0 \left( \frac{E}{E_b} \right)^{-(\alpha + \beta(\log(E/E_b))},
\end{equation}

\noindent where $E_b$ was fixed to the FL8Y catalog value of 457.698 MeV.

%% power law
% \begin{equation}\label{eq:pl}
%     \frac{dN}{dE} = N_0 \left( \frac{E}{E_0} \right)^{-\gamma}
% \end{equation}

%% plSEC
% \begin{equation}
%     \frac{dN}{dE} = N_0 \left( \frac{E}{E_0} \right)^{-\gamma_1} e^{-(E/E_c)^{-\gamma_2}}
% \end{equation}

We checked that the log-parabola model provides a better fit than a power-law model using the likelihood ratio test. A power-law sub-exponential cutoff model was also preferred over a power law, but we could not establish whether this model is significantly preferred with respect to the log-parabola model using a likelihood ratio test. This is because the two curved models are non-nested, i.e. neither is a special case of the other, and therefore it is not possible to calculate the statistical significance of a preference for one over the other. We assumed a log-parabola spectrum for all subsequent LAT analyses. To facilitate comparison with the VERITAS points, the LAT model fits and butterfly contours were extended beyond the LAT maximum energy of 500 GeV, and extragalactic background light absorption was applied to them using the model of \cite{Franceschini_2017}.

The global spectral shapes of the three sources are similar, with an index $\alpha$ of ${\sim}$2.1--2.3 and a curvature parameter $\beta$ of ${\sim}$0.04--0.06, and they differ primarily by their normalization.

\begin{figure*}[htbp]
\includegraphics[width=\textwidth]{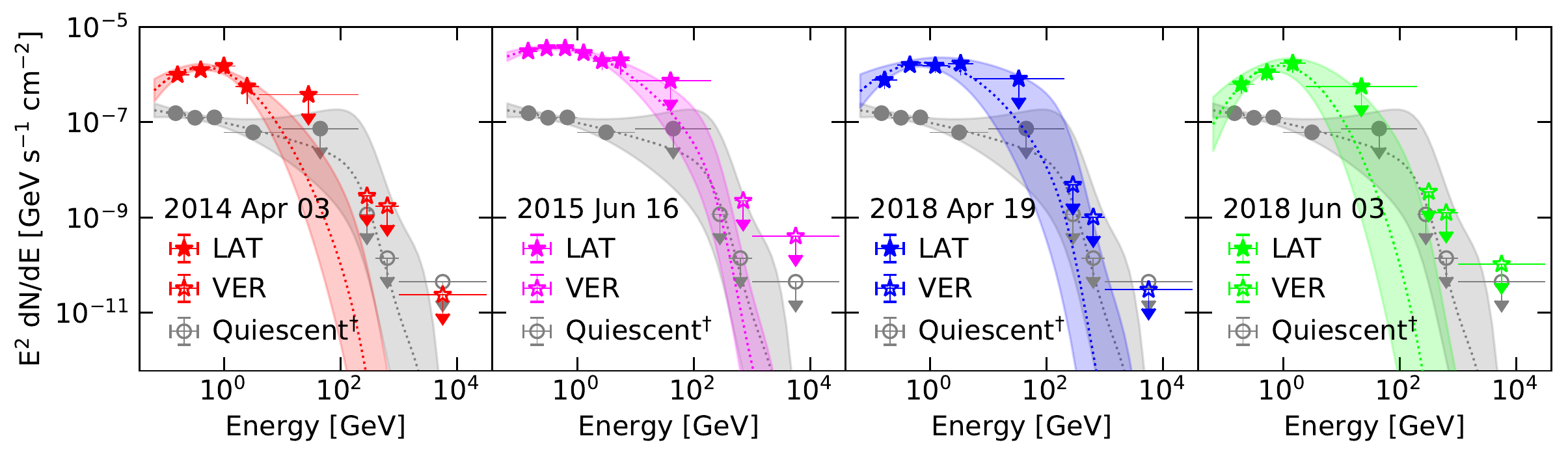}
\caption{\textit{Fermi}-LAT spectrum and VERITAS spectral upper limits of 3C 279 during four flares, strictly simultaneous with the VERITAS observations. The quiescent spectrum (gray circles and contour) is shown for comparison. The strictly simultaneous quiescent state LAT data and VERITAS upper limits are represented with filled and unfilled markers, respectively. The LAT spectra are extrapolated to the VERITAS energy range, incorporating EBL absorption. Downward arrows show 95\% confidence level upper  limits.}
\label{fig:veritas_seds}
\end{figure*}

Using the data from 3C~279, we compared several methods to determine a baseline non-flaring spectrum. First, we defined a low state lasting from MJD 56230 to 56465 (see Figure \ref{fig:graylcs}), during which the flux was quiescent and stable in HE gamma rays, $R$-band optical, and X-rays. We checked publicly available Tuorla\footnote{\url{https://users.utu.fi/kani/1m/3C_279_jy.html}} data for the $R$-band light curve. For the X-rays, we analyzed the \textit{Swift}-XRT light curve using the online data products generator\footnote{\url{https://www.swift.ac.uk/user_objects/}}. To ensure low, stable gamma-ray emission, we selected the interval to span the Bayesian blocks with the lowest flux while excluding intervals with the Sun in the ROI. The low-state LAT SED is shown in Figure~\ref{fig:global_seds}. Only one VERITAS observation occurred during this interval, on MJD 56417, so the corresponding VERITAS upper limits are not constraining and are not shown.

\begin{figure*}[htbp]
\includegraphics[width=\textwidth]{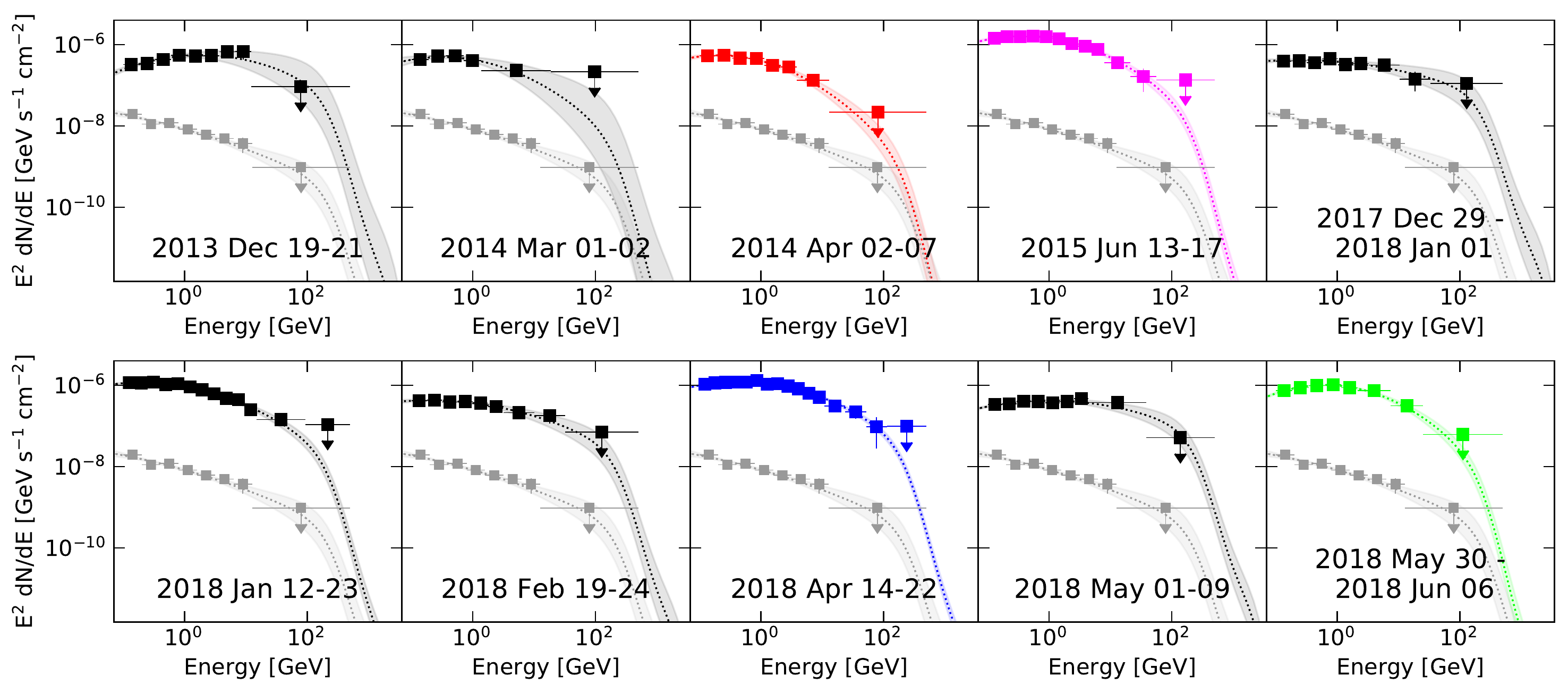}
\caption{
\textit{Fermi}-LAT spectra of 3C 279 during ten flares, for the intervals selected using the algorithm proposed in this work and described in Section \ref{sec:flarealgorithm}. For comparison, the LAT low state spectrum is shown in gray squares in all of the panels. The four flares shown in color have a corresponding spectrum in Figure~\ref{fig:veritas_seds}.}
\label{fig:flare_seds}
\end{figure*}

Next, using the algorithm proposed in this work and described in Section \ref{sec:flarealgorithm}, we designated all epochs of the LAT light curve other than the flaring episodes as quiescent. From those epochs, we extracted those LAT data strictly simultaneous with the VERITAS observations, integrating a total of 43.6 hours of observations. The resulting strictly simultaneous LAT spectrum and VERITAS spectral upper limits are shown in Figure~\ref{fig:global_seds}. We then performed the same procedure for four flaring epochs during which a significant \textit{Fermi}-LAT detection could be obtained strictly simultaneous with the VERITAS observations, which occurred on the nights of April 3, 2014; June 16, 2015; April 19, 2018; and June 3, 2018. These strictly simultaneous LAT and VERITAS SEDs are shown in Figure~\ref{fig:veritas_seds}.

The spectral shapes of the 3C~279 low and quiescent states are similar to each other and to the global state, although the uncertainties on their fit parameters are high due to the low significance. The spectra differ primarily in their flux normalization. The normalization of the low state is lower than that of the global state by design, while the normalization of the strictly simultaneous quiescent state is higher. This could result from the timing of the VERITAS monitoring and triggered observations which often follow up on \textit{Fermi}-LAT flares and may tend to catch mildly elevated activity in \textit{Fermi}-LAT even if the source is not actually flaring.

% Show a quiescent SED including all of the data in the quiescent state, not just the strictly simultaneous points. This would confirm that the method makes sense, and demonstrate the power of the method to expand the usable LAT statistics.

Finally, we derived LAT SEDs for all of the ten identified flares of 3C~279, using the entire flare time periods, irrespective of strict simultaneity with VERITAS, shown in Figure~\ref{fig:flare_seds}. The average flare spectrum is more strongly curved than the global spectrum, with $\alpha = 2.02 \pm 0.01$ and $\beta = 0.093 \pm 0.008$, compared to $\alpha = 2.228 \pm 0.004$ and $\beta = 0.061 \pm 0.003$ for the global state.

%What do we learn from spectra? What can their log-parabolic shape after factoring out the EBL tell us about primary particle energies? What are the implications of the TeV detections of PKS~1222+216 and Ton~599?

\section{SED Modeling}\label{sec:sed_modeling}

Multiwavelength SED modeling can shed light on the mechanisms of gamma-ray production during VHE flares. For 3C~279, we refer the reader to those works in the literature in which multiwavelength SED modeling of the epochs considered here has been performed, and we do not perform any additional modeling \citep[see for example,][]{Hayashida2015, Ackermann2016, Prince2020, Yoo2020}.

PKS~1222+216 was first detected at TeV energies by MAGIC during a flaring event in June 2010 \citep{Aleksic2011}, and multiwavelength SED modeling of this event has been performed by e.g. \cite{Tavecchio2011}. We therefore restricted our SED modeling of the source to the duration of the second VHE detection by VERITAS in February and March 2014. We considered data from all instruments taken from UT 2014-02-26 to 2014-03-10, inclusive.

Ton~599 has not been studied as extensively as the other two sources. \cite{Prince2018} and \cite{Patel2020} model its variability characteristics and multiwavelength SED, respectively, during the high state in December 2017, but do not have access to TeV data. We therefore modeled the multiwavelength SED of Ton 599 during the VERITAS detection in December 2017. We considered data from all instruments taken from UT 2017-12-15 to 2017-12-16, inclusive.

To assemble our multiwavelength SEDs, in addition to the gamma-ray data from VERITAS and \textit{Fermi}-LAT, we incorporated X-ray and ultraviolet data from the XRT and UVOT instruments aboard the \textit{Swift} satellite and optical observations from the Steward Observatory.

\begin{figure}
    \includegraphics[width=0.45\textwidth]{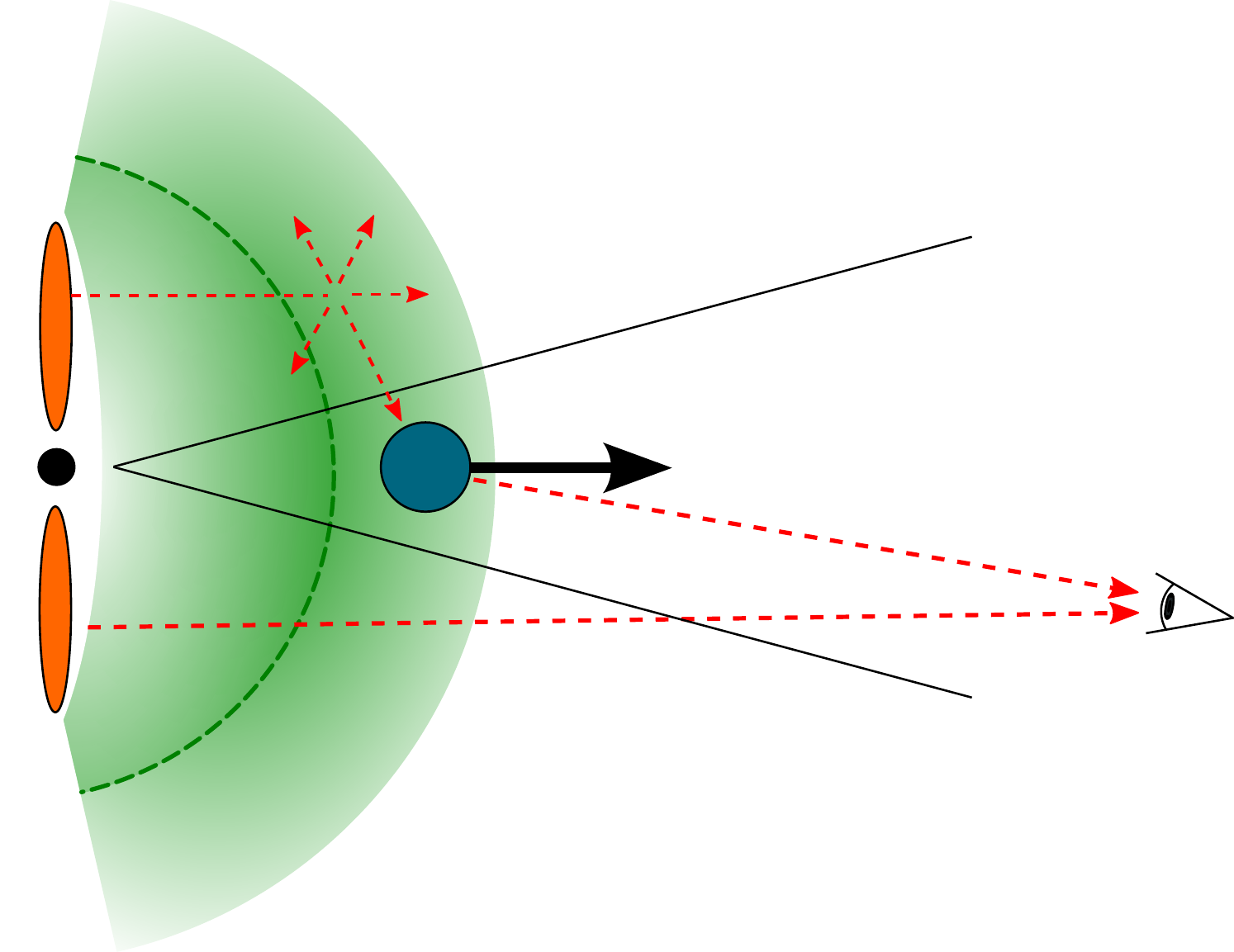}
    \put(-215,88){\colorbox{white}{\bf \scriptsize SMBH}}
	\put(-235,35){\colorbox{white}{\bf \scriptsize Accretion disk}}
	\put(-185,160){\colorbox{white}{\bf \scriptsize BLR}}
	\put(-165,70){\colorbox{white}{\bf \scriptsize Blob}}
	\put(-85,125){\colorbox{white}{\bf \scriptsize Jet}}
	\put(-42,48){\colorbox{white}{\bf \scriptsize Observer}}
    \caption{A schematic illustration of the emission model used in this work (not to scale). %The central supermassive black hole (black circle) is surrounded by the accretion disk (orange regions), with the BLR (green region) farther out. The BLR reprocesses emission from the accretion disk and has a density decreasing with distance from the SMBH as 1/$r^2$. The gamma-ray emission zone (blue circle) lies inside the jet (black lines) and interacts with emission from the BLR. 
    The green dashed arc represents the nominal BLR radius $r_\mathrm{BLR}$ corresponding to the region of the maximal BLR density.
    The observer measures the beamed emission from the blob interacting with the BLR as well as the accretion disk's thermal emission.
    %We note that no spatial extension of the accretion disk is considered in our model.
    The accretion disk is assumed to be a point source.}
    \label{fig:sed_model_illustration}
\end{figure}
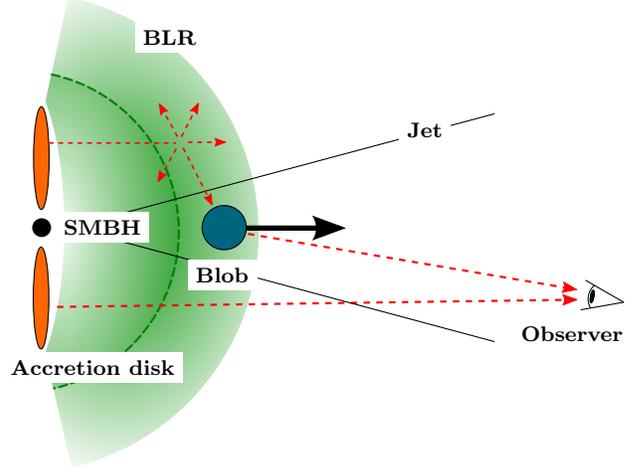

We described the multiwavelength SEDs of the two FSRQs using a multi-component synchrotron self-Compton (SSC) blob-in-jet model, implemented using the framework of the ``\texttt{Bjet}" code, developed by \cite{Hervet2015} and based on \cite{Katarzynski2001}. We modeled the radiative interactions of a compact leptonic emission zone (a blob), including an EIC emission component resulting from the interactions of the blob particles with the thermal accretion disk emission reprocessed by the BLR. Figure~\ref{fig:sed_model_illustration} shows a schematic illustration of the components producing the emission in this model.

%In our model, we considered the blob position to be stationary in the rest frame of the galaxy. This assumption can be interpreted as a relativistic particle flow crossing a standing shock, with the blob having the typical size of the shock cross section.
%, with the blob radius representing the distance from the shock in the jet frame required for the particles to cool, which is roughly equal to the size of the shock if the cooling time is shorter than the shock-crossing time. 
%However, we note that in a stationary SSC model this scenario is formally indistinguishable from that of a relativistically moving blob.
%OH: the upper description can be tricky since the size of the blob would be wavelength dependant in the case of a continuous flow crossing a standing shock. Probably better to remove this part since it doesn't have any implications in the following discussions-conclusions

We consider a simplified BLR model with a normalized density profile, based on \cite{Nalewajko2014}, where $\rho_\mathrm{BLR}(r)$ is at a maximum at the characteristic BLR radius $r = r_\mathrm{BLR}$ and decreasing as $r^{-2}$ with the distance to the core such that
\begin{equation}
    \rho_\mathrm{BLR}(r) = \frac{(r/r_\mathrm{BLR})^2}{1+(r/r_\mathrm{BLR})^4}~,
\end{equation}
with $r_\mathrm{BLR}$ scaled to the bolometric disk luminosity $L_d$ as $r_\mathrm{BLR} = 0.1 \sqrt{L_d/1\times10^{46}~\mathrm{erg}~\mathrm{s}^{-1}}$~pc \citep{Sikora2009, Ghisellini2009}. From SED modeling of PKS~1222+216 and Ton~599 we deduce a BLR radius of 0.17 pc and 0.15 pc respectively.
We assume an isotropic diffusion of the disk light by the BLR, where the specific intensity of this field can be expressed as 
\begin{equation}
I_\mathrm{BLR}(\nu, T_{d}, r) = \epsilon_\mathrm{BLR} \rho_\mathrm{BLR}(r) \frac{L_{d}}{4\pi r^2}\frac{I_{p}(\nu, T_{d})}{(\sigma_\mathrm{SB}/\pi)T_{d}^4}
,\end{equation}
where $\sigma_\mathrm{SB}$ is the Stefan-Boltzmann constant, $I_{p}$ is the Planck intensity, and $\epsilon_\mathrm{BLR}$ is the covering factor. This equation is similar to Eq. 12 in \cite{Hervet2015} with the addition of the BLR density profile.
Only the extension of the BLR in front of the blob plays a significant role in our modeling since it drives the number of gamma rays produced by the blob that will be absorbed by pair production. 
The BLR is by default defined between $r = 0$ and $r = 100~r_\mathrm{BLR}$. Given the fast convergence of the BLR opacity ($I_\mathrm{BLR} \propto r^{-4}$), the maximum extension of the BLR does not play a significant role in the model. Although we assume for simplicity that the BLR is isotropic, any anisotropy should have a small effect on the opacity \citep[e.g.][Figure 14]{Abolmasov2017}.

Figures~\ref{fig:pks1222_mwlsed} and \ref{fig:ton599_mwlsed} show the multiwavelength SED models of PKS~1222+216 and Ton~599. In these figures, the synchrotron and SSC emission are shown by solid blue lines. The subdominant second-order self-Compton emission caused by the interactions of the electrons with the self-Compton photons is shown by a dotted blue line. The thermal emission from the accretion disk is modeled as a point source radiating as a black body, and is shown by a heavy dashed green line. The inverse Compton emission due to the interaction of the electrons with the disk photons reprocessed in the BLR is shown by a dashed green line. Table~\ref{tab:sed_model_params} gives the parameters characterizing the SED models.

Our model does not include any secondary radiation from pair cascades produced by the absorption of gamma rays in the BLR. While detailed modeling of this effect is beyond the scope of this paper, we estimate that the potential contribution of such cascades would be $\ll 1\%$ of the total bolometric luminosity for PKS~1222+216 and $\lesssim 1\%$ for Ton~599, given the respective levels of absorption in our models, which are described below. We evaluated these contributions by comparing the radiative output of each model with that from the same model with the BLR opacity set to zero. This effect may be noted as a source of systematic uncertainty when interpreting our results.

We note that our model requires that the dust-torus luminosity be negligible compared to the disk luminosity. As evidence of far-infrared dust-torus thermal emission is lacking in the SED, we consider this assumption to be reasonable in our study. Observing campaigns with good microwave to IR coverage would be needed to fully confirm this approach. The presence of strong dust-torus emission would require that the gamma-ray emission zone be farther downstream in the jet so as not to produce too large an opacity by pair production.

\subsection{PKS 1222+216 modeling}

\begin{figure*}
    \includegraphics[width=0.5\textwidth]{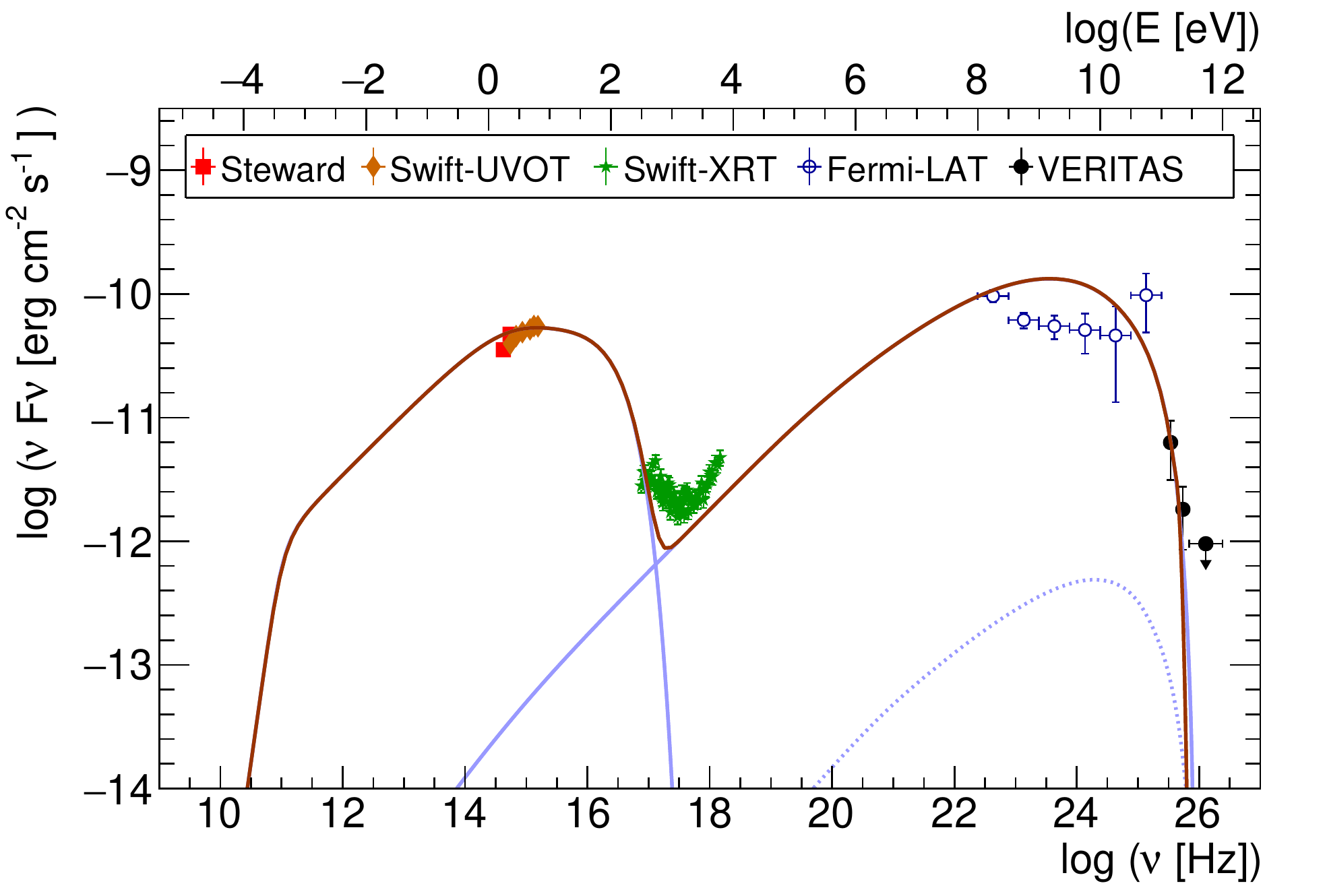}
    \includegraphics[width=0.5\textwidth]{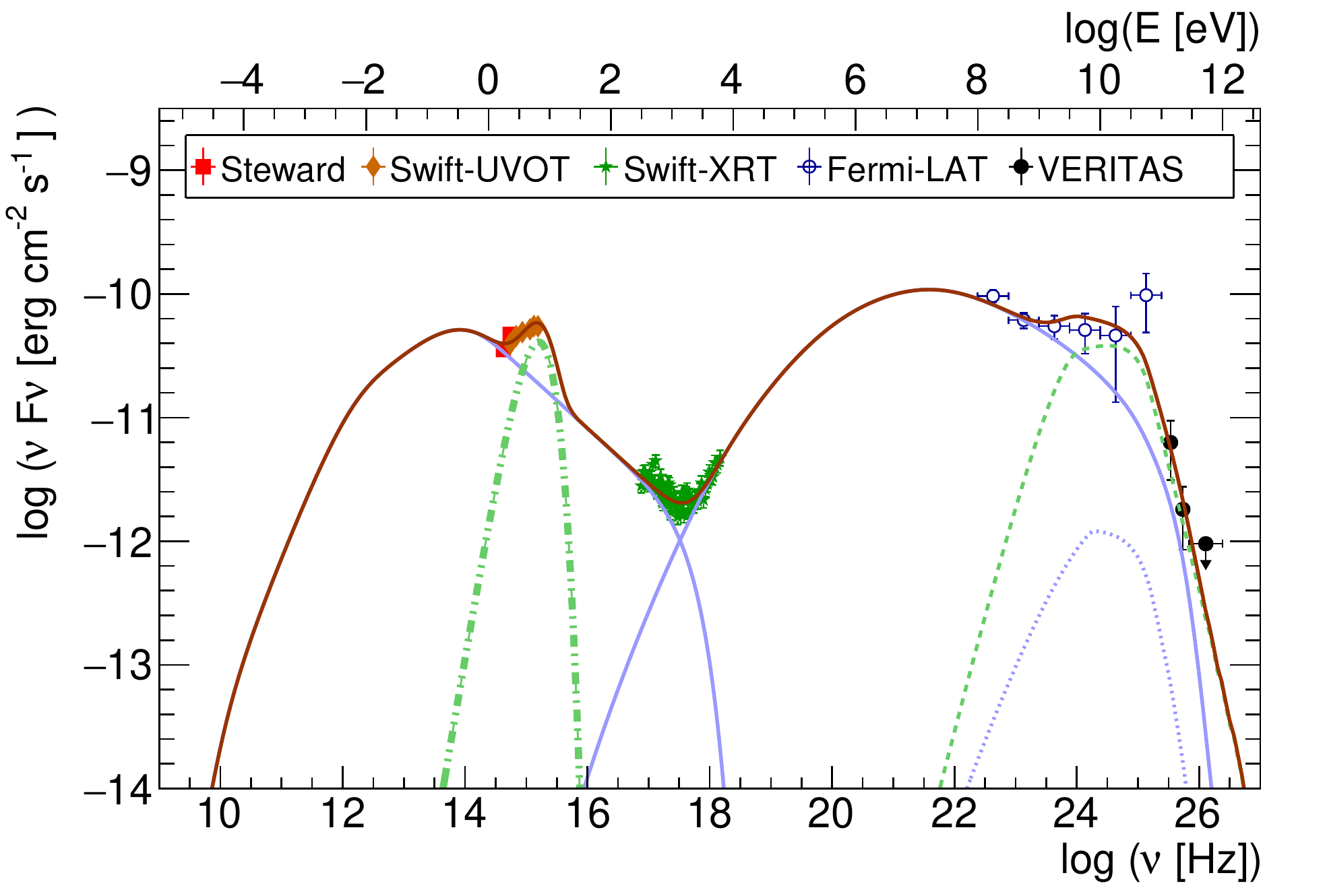}
    \caption{Broadband SED of PKS~1222+216 during the VERITAS detection from UT 2014-02-26 to 2014-03-10. \textit{Left:} Pure synchrotron self-Compton model. \textit{Right:} Model considering an external inverse Compton component at high energy from the interaction of blob particles with the thermal accretion disk emission reprocessed by the BLR. The solid blue lines show synchrotron and SSC emission; the dotted blue line shows second-order self-Compton emission; the heavy dashed green line shows thermal emission from the accretion disk; and the dashed green line shows inverse Compton emission from the BLR. The EBL absorption is taken into account considering the model of \protect{\cite{Franceschini_2017}}.}
    \label{fig:pks1222_mwlsed}
\end{figure*}

In order to investigate the necessity of including an EIC component, we represented the multiwavelength SED of PKS~1222+216 with a one-zone pure SSC model, shown in Figure~\ref{fig:pks1222_mwlsed} (\textit{left}). As can be seen by the similar amplitudes of the synchrotron and inverse Compton peaks in the figure, the SED is only weakly Compton dominated, with the inverse Compton luminosity about twice the synchrotron luminosity. The \textit{Swift}-XRT spectrum contains a well-resolved break showing the transition between synchrotron and inverse Compton dominated emission, which sets a strong constraint on the model.
Our best attempt does not provide a satisfying representation of the observed SED. The main issue is that the optical-to-X-ray components of the SED have steep slopes which would require a narrow, sharp synchrotron bump to achieve a good representation, while the X-ray-to-VHE needs a wide, flat inverse Compton bump. This is not compatible with the usual simple SSC framework, especially when the SED is not heavily Compton dominated.

%The hard X-ray spectral index of .... and the relatively flat spectrum of Fermi-LAT do not agree with a single spectral component, since it would induce seveal breaks in the SED not consistent with an SSC emission.
%The visible Fermi excess at low energy also induces that it belongs to a different radiative origin than the VHE energy part.

In our EIC model, the IR-to-UV SED is dominated by the blackbody big-blue-bump emission of an accretion disk (see Figure~\ref{fig:pks1222_mwlsed}, \textit{right}), which resolves the tension by eliminating the constraint on the synchrotron spectral shape. This allows for a broad SSC component matching the spectral break observed in the X-ray band. In this scenario, the VHE emission is produced by the EIC process between a relativistic blob and the disk thermal emission reprocessed by the BLR. The blob is set to a distance of 3.56 pc from the SMBH, corresponding to $21.3~r_\mathrm{BLR}$.
It should be noted that a thermal EIC process was also favored in previous models of PKS~1222+216 where clear disk emission and a strongly Compton-dominated SED were observed during a major outburst in 2010 \citep{Tavecchio2011}.

Because the peak frequency of the EIC emission is directly proportional to the blob Lorentz factor, this scenario imposes a strong constraint on the jet parameters. For PKS~1222+216, in order to match the VHE spectrum, the bulk Lorentz factor needs to be above approximately 23, which was achieved by assuming a Doppler factor $\delta = 40$ and an angle with the line of sight $\theta_\mathrm{obs} = 1\degree$.
%This assumption is consistent with the parameter space presented in Figure~\ref{fig:nalewajko_constraints} where a Lorentz factor of $\Gamma = 23$ would lead to an emission zone located within a distance of $\sim 0.3-10$ pc of the SMBH. 
This assumption is consistent with the jet constraints derived by \citet{Hervet_2016} from the fastest motion observed in the radio jet of PKS~1222+216, which led to estimations of $\theta_\mathrm{obs} = 1.3\degree$, $\delta = 41.3$ and $\Gamma = 29.2$.

%new text
Because no significant variability was observed in any waveband during the time period selected for modeling for either source, we considered a stationary model giving a snapshot of the observed activity.
As a consistency check, we compared the expected radiative cooling time from the model with the observed flare decay timescale. The cooling time associated with the full radiative output (synchrotron, SSC and EIC emissions) can be expressed in the Thomson regime as 
\begin{equation}
T_{\mathrm{cool}} (\gamma) = \frac{3 m_e c}{4 \sigma_T \gamma  (U'_B + U'_\mathrm{syn} + U'_\mathrm{blr})},
\end{equation}
with $m_e$ the electron mass, $\sigma_T$ the Thomson cross section, $\gamma$ the Lorentz factor of the emitting particle, and  $U'_B$, $U'_\mathrm{syn}$, $U'_\mathrm{blr}$ respectively the energy density in the blob frame of the magnetic field, synchrotron field, and external BLR field \citep[e.g.][]{Inoue_1996}.
One can associate the energy at the break of the spectral particle distribution $\gamma_{\mathrm{brk}}$ with the emission at the peaks of the SED. The \textit{Fermi}-LAT energy range being mostly above this peak, we can deduce $T_\mathrm{cool}(\mathrm{\textit{Fermi}}) \lesssim 17$ days. This is consistent with the observed \textit{Fermi}-LAT flare decay of $10.4\pm6.2$ days.

The minimum possible variability predicted by our model is 18 h, given by the blob's radius and Doppler factor such that $\tau_{\mathrm{min}} = R (1+z)/(c \delta)$.
%old text
%Our model predicts a possible minimal variability of 18 hours, relatively close to the fastest variability observed during the full period of GeV flaring activity. 
The total power of the jet is approximately $3.4\times 10^{45}$ erg s$^{-1}$, in a particle-dominated regime with the equipartition parameter $U_B/U_e =1.7\times 10^{-3}$.

\subsection{Ton~599 modeling}

\begin{figure}
    \includegraphics[width=0.5\textwidth]{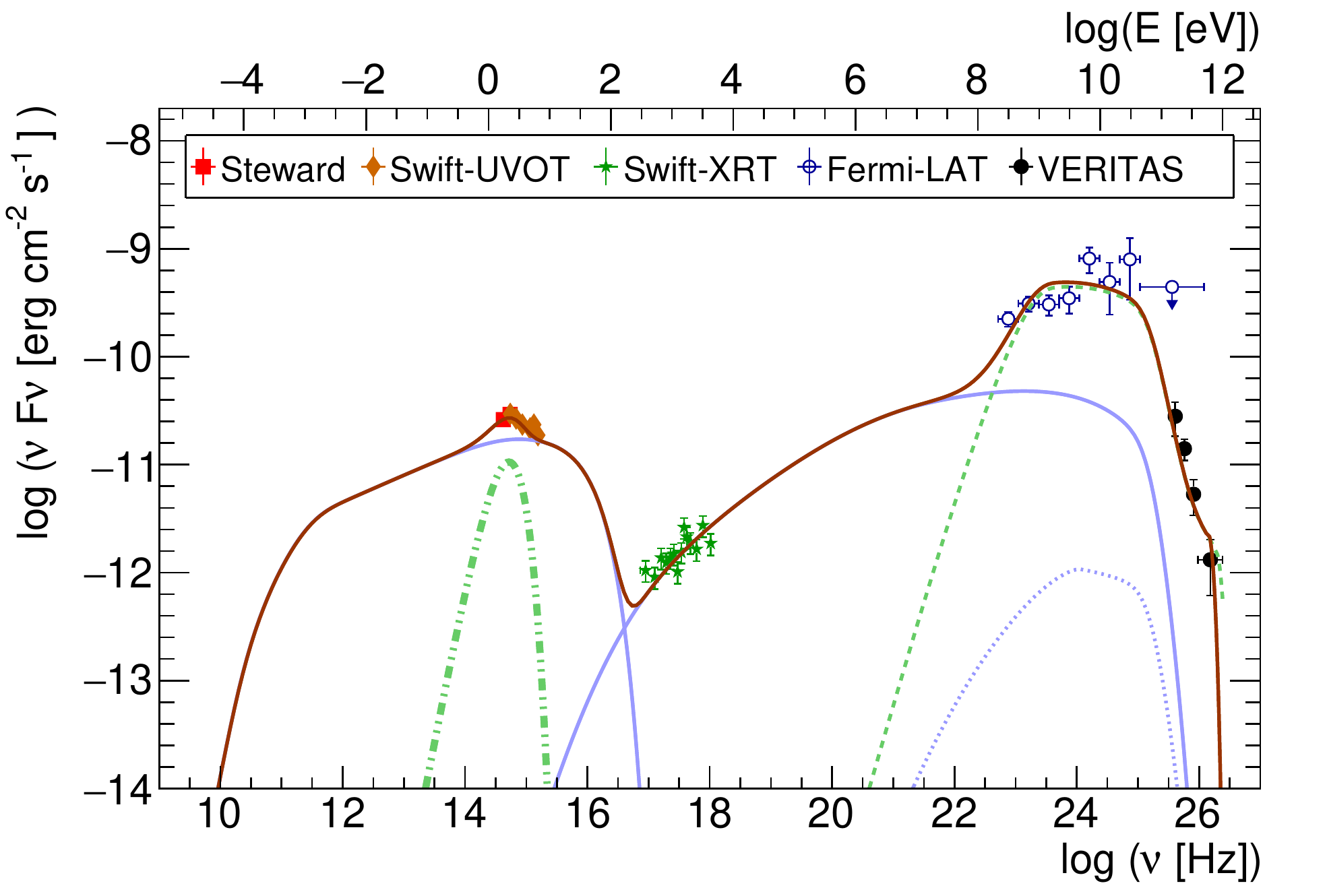}
    \caption{Broadband SED of Ton~599 during the VERITAS detection from UT 2017-12-15 to 2017-12-16. The solid blue lines show synchrotron and SSC emission; the dotted blue line shows second-order self-Compton emission; the heavy dashed green line shows thermal emission from the accretion disk; and the dashed green line shows inverse Compton emission from the BLR.}
    \label{fig:ton599_mwlsed}
\end{figure}

Contrary to PKS~1222+216, the SED of Ton~599 is heavily Compton dominated, with a ratio of inverse Compton to synchrotron luminosity of approximately one order of magnitude. This is a usual signature of an EIC component dominating the gamma-ray emission. We therefore consider the same scenario as for PKS~1222+216. As shown in Figure~\ref{fig:ton599_mwlsed}, the model describes the data well.

As in the case of PKS 1222+216, the thermal EIC emission imposes strong constraints on the properties of the emitting region. The largest constraint comes from the gamma-ray opacity by pair creation from the luminous thermal field surrounding the blob. We found that only for a Doppler factor of ${\gtrsim} 50$ is the EIC emission at VHE strong enough to produce the observed VHE gamma rays, given the BLR opacity.

\begin{deluxetable}{cccc}[ht!]
\tablecaption{Parameters of the SED models.\label{tab:sed_model_params}}
\tabletypesize{\scriptsize}
    \tablehead{
    \colhead{Parameter} & \colhead{PKS~1222+216} & \colhead{Ton~599} & \colhead{Unit}\\ \hline
    \colhead{$\theta_\mathrm{obs}$} & \colhead{$1.0$} & \colhead{$1.0$} &  \colhead{deg}\\ \hline
    \colhead{Blob} &  &
    }
    \startdata
    $\delta$ & $40$ & $53$\\ 
    $N_{e}^{(1)}$ & $2.0\times 10^{4}$ & $2.7\times 10^{5}$ & cm$^{-3}$\\ 
    $n_1$ & $2.1$ & $2.5$ & $-$ \\
    $n_2$ & $3.9$ & $3.0$ & $-$\\
    $\gamma_{\mathrm{min}}$ & $5.5\times 10^{2}$ & $3.0\times 10^{2}$ & $-$\\
    $\gamma_{\mathrm{max}}$ & $3.0\times 10^{5}$ & $7.0\times 10^{4}$ & $-$\\
    $\gamma_{\mathrm{brk}}$ & $5.0\times 10^{3}$ & $1.5\times 10^{4}$ & $-$\\
    $B$ & $3.0\times 10^{-2}$ & $3.0\times 10^{-2}$ & G\\
    $R$ & $5.5\times 10^{16}$ & $6.0\times 10^{16}$ & cm\\
    $D_\mathrm{BH}$\tablenotemark{*} & $3.56$ & $2.33$ & pc\\
    \hline
    Nucleus\\
    \hline
    $L_\mathrm{disk}$ & $2.8\times 10^{46}$ & $2.2\times 10^{46}$ & erg s$^{-1}$\\ 
    $T_\mathrm{disk}$ & $2.8\times 10^{4}$  & $1.1\times 10^{4}$ & K\\
    $\epsilon_\mathrm{BLR}$ & $2.0\times 10^{-2}$ & $2.0\times 10^{-2}$ & $-$\\
\enddata
\tablecomments{$\theta_\mathrm{obs}$ is the angle of the blob direction of motion with respect to the line of sight. The electron energy distribution between Lorentz factors $\gamma_\mathrm{min}$ and $\gamma_\mathrm{max}$ is given by a broken power law with indices $n_1$ and $n_2$ below and above $\gamma_\mathrm{brk}$, with $N_e^{(1)}$ the normalization factor at $\gamma = 1$. The blob Doppler factor, magnetic field, radius, and distance to the black hole are given by $\delta$, $B$, $R$, and $D_\mathrm{BH}$, respectively. The disk luminosity and temperature are given by $L_\mathrm{disk}$ and $T_\mathrm{disk}$, while $\epsilon_\mathrm{BLR}$ is the covering factor of the broad line region.}
\tablenotetext{*}{\textit{ Host galaxy frame.}}
\end{deluxetable}

The solution presented in Figure~\ref{fig:ton599_mwlsed}, with $\delta = 53$, is consistent with a maximum VHE emission undergoing strong BLR absorption ($E_\mathrm{max} = 630$ GeV), with an opacity of $\tau_{\gamma \gamma,E_\mathrm{max}} = 2.8$. In this scenario we set the blob at a distance of 2.33 pc from the SMBH, corresponding to $15.7~r_\mathrm{BLR}$.

% new text
By applying the same consistency check for variability as PKS~1222+216, we found a minimal possible variability timescale predicted by the model of 18 h (coincidentally the same as PKS~1222+216), and a radiative cooling time 
$T_\mathrm{cool}(\mathrm{\textit{Fermi}}) \simeq 8.7$ days, in good agreement with the observed \textit{Fermi}-LAT flare decay of $11.8\pm1.1$ days.
The VERITAS observed variability can be associated with the cooling time $T_\mathrm{cool}(\gamma_{\mathrm{max}})$, which leads to $T_\mathrm{cool}(\mathrm{VERITAS}) = 45$ h, fully compatible with the observed variability of $\sim$ 2 days.
%old text
%By coincidence, the fastest possible variability of Ton~599 from the model is 18 hours, similar to that of PKS~1222+216, consistent with the variability timescale of ${\sim}2$ days observed with VERITAS. 
The blob is estimated to have a total power of approximately $1.2\times 10^{46}$ erg s$^{-1}$, and to be extremely particle-dominated with equipartition parameter $U_B/U_e =3.8\times 10^{-4}$.

\begin{deluxetable*}{lccccccccccc}
\tablecaption{Parameters used to calculate constraints on the parameter space.\label{tab:constraint_parameters}}
%\tabletypesize{\scriptsize}
\tablehead{
\colhead{Source} & \colhead{z} &  \colhead{$D_L$} & \colhead{$t_\mathrm{var}$}  & \colhead{$L_\mathrm{syn}$} & \colhead{$L_\mathrm{gamma}$} & \colhead{$L_d$} & \colhead{$M_\mathrm{BH}$\tablenotemark{a}} & \colhead{$E_\mathrm{cool}$} & \colhead{$\epsilon_\mathrm{BLR}$\tablenotemark{b}} & \colhead{$\epsilon_\mathrm{IR}$\tablenotemark{b}}\\
\colhead{} & \colhead{} &  \colhead{Gpc} & \colhead{day}  & \colhead{erg s$^{-1}$} & \colhead{erg s$^{-1}$} & \colhead{erg s$^{-1}$} & \colhead{$M_\odot$} & \colhead{GeV} & \colhead{} & \colhead{}
}
\startdata
PKS 1222+216 & 0.434 & 2.44 & 10.0 & $3.5 \times 10^{46}$ & $7.8 \times 10^{46}$ & $2.8 \times 10^{46}$ & $3.47 \times 10^8$ & 7.07 & 0.02 & 0.2\\
Ton 599 & 0.725 & 4.54 & 2.0 & $4.4 \times 10^{46}$ & $1.2 \times 10^{48}$ & $2.2 \times 10^{46}$ & $6.8 \times 10^8$ & 326 & 0.02 & 0.2
\enddata
\tablecomments{$z$ and $D_L$ are the redshift and luminosity distance of the source. $t_\mathrm{var}$ is the variability timescale of cooling derived from each flare's fitted exponential decay. $L_\mathrm{syn}$, $L_\mathrm{gamma}$, and $L_d$ are the synchrotron luminosity, gamma-ray luminosity, and disk luminosity from the SED model. $M_\mathrm{BH}$ is the black hole mass. $E_\mathrm{cool}$ is the maximum photon energy due to the external Compton cooling of relativistic electrons. $\epsilon_\mathrm{BLR}$ and $\epsilon_\mathrm{IR}$ are the covering factors of the broad line region and IR-emitting torus region, respectively.}
\tablenotetext{a}{\citet{Farina2012, Liu2006}}
\tablenotetext{b}{\citet{Tavecchio2011}}
\end{deluxetable*}

\section{Lorentz factors and locations of the gamma-ray-emitting regions}\label{sec:discussion_emission_region}

\begin{figure*}
    \includegraphics[width=0.5\textwidth]{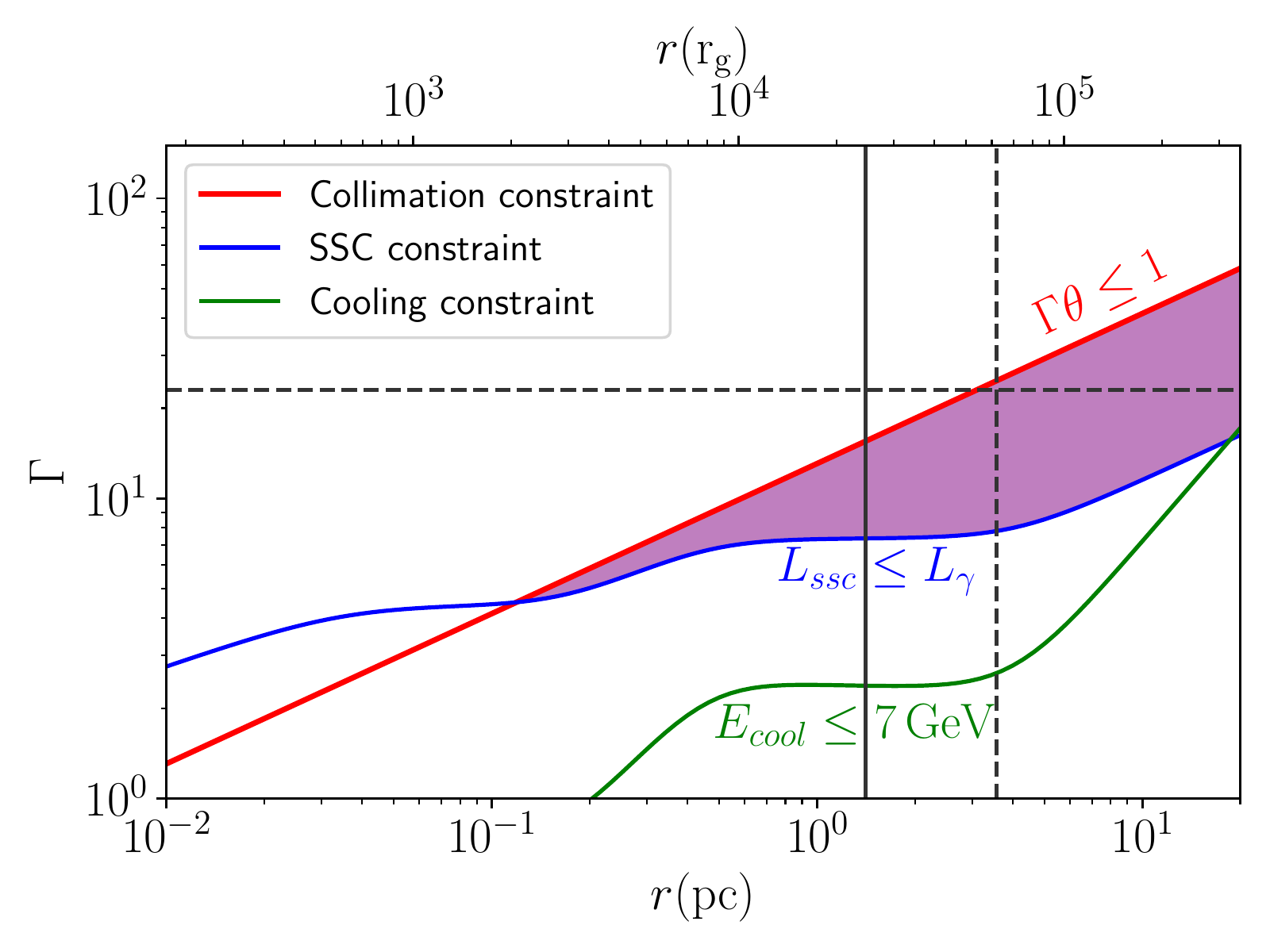}
    \put(-155,190){\makebox(0,0)[lb]{\textbf{PKS~1222+216}}}
    \includegraphics[width=0.5\textwidth]{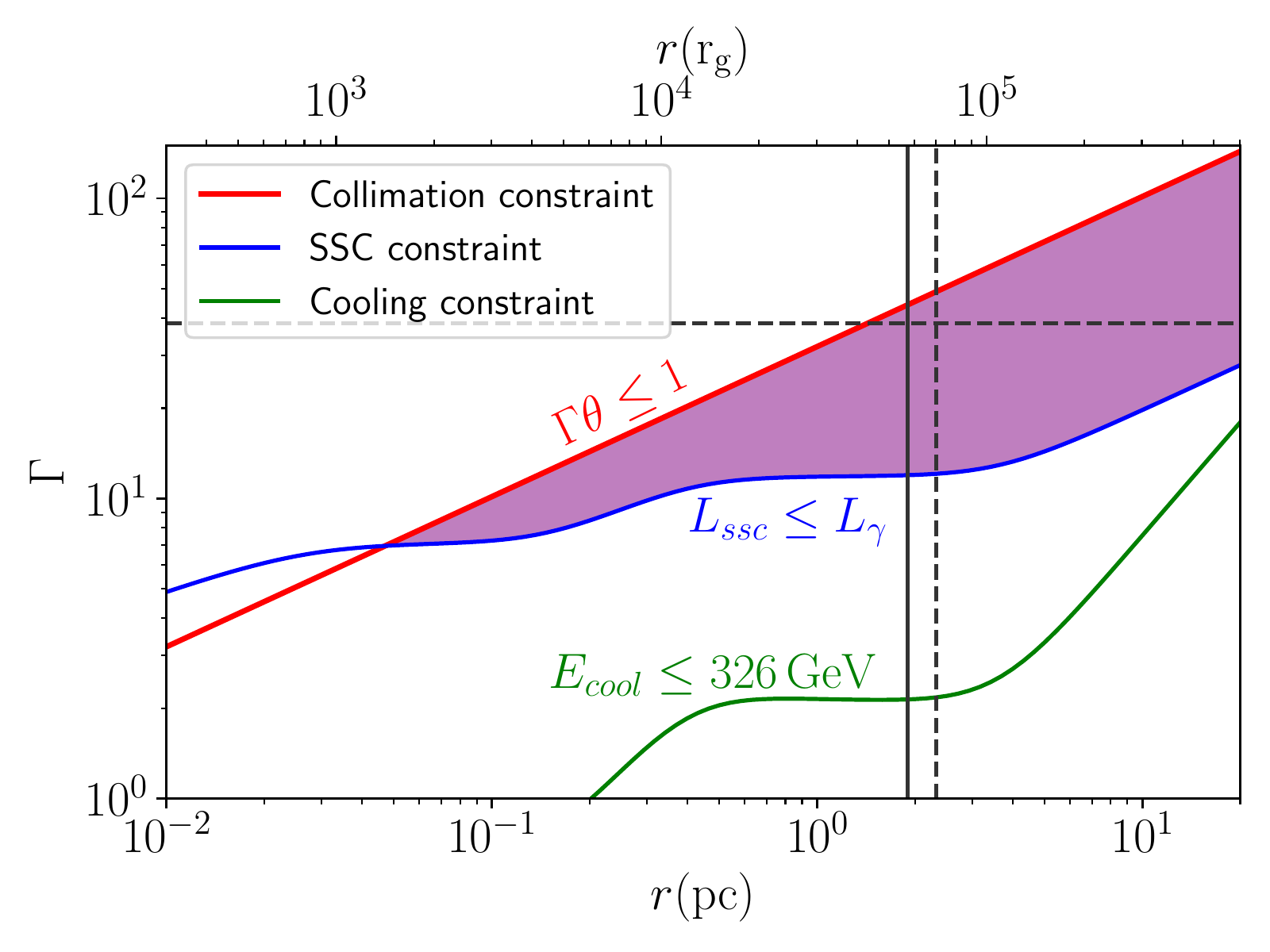}
    \put(-140,190){\makebox(0,0)[lb]{\textbf{Ton~599}}}
    \caption{Constraints on the Lorentz factor $\Gamma$ and distance $r$ between the gamma-ray emission location and central black hole, adapted from \cite{Nalewajko2014}. The allowed region is filled in purple. The black vertical line shows the opacity constraint on $r$ derived from the BLR modeling. The values of $\Gamma$ and $r$ derived from the SED modeling are indicated with dashed black lines.}
    \label{fig:nalewajko_constraints}
\end{figure*}

We determined constraints on the Lorentz factor $\Gamma$ and distance $r$ of the gamma-ray emission region from the central black hole for PKS~1222+216 and Ton~599 following the method and assumptions of \citet{Nalewajko2014}. The constraints are plotted in Figure~\ref{fig:nalewajko_constraints}. The parameters used to determine the constraints are given in Table~\ref{tab:constraint_parameters}. In order to obtain a conservative SSC constraint, we set the SSC luminosity equal to the observed gamma-ray luminosity $L_\mathrm{gamma}$. For PKS~1222+216 the fastest variability is observed with \textit{Fermi}-LAT, while for Ton~599 it is observed with VERITAS. We therefore set the maximum energy $E_\mathrm{cool}$ for the EIC cooling constraint equal to the geometric mean of the energy ranges observed by \textit{Fermi}-LAT and VERITAS for the two sources, respectively.

Three constraints on $\Gamma$ and $r$ are calculated. The collimation constraint requires that the size of the emission region be less than the size of the jet at the emission region location such that $\Gamma \theta \leq 1$, where $\theta$ as defined by \citet[][Eq.~1]{Nalewajko2014} is the angle subtended by the blob expanding while propagating. %In our case, considering a stationary shock, it would refer to the expansion of the relativistic flow passing through the shock.
%opening angle of emission from the blob
%jet opening angle determined by the blob size

A caveat of the collimation constraint is the underlying assumption that the blob size is defined by the observed variability such that $R = c \delta t_\mathrm{var,obs} /(1+z)$. However, the observed variability gives only an upper limit on the blob radius, meaning that the actual size of the emission zone is likely smaller than that extrapolated from the observed variability. Indeed, our modeling of PKS~1222+216 and Ton~599 predicts observed variability much shorter than the one observed within the reconstructed SED periods. This discrepancy explains why the parameters predicted by our model lie only just below the line $\Gamma \theta = 1$.

The SSC constraint requires that the SSC luminosity should not exceed the total gamma-ray luminosity, which includes contributions from external radiation fields \citep[][Eq.~5]{Nalewajko2014}. The cooling constraint requires that electrons radiatively emitting gamma rays at energies above $E_\mathrm{cool}$ cool through interactions with external radiation fields faster than the flare decay timescale \citep[][Eq.~9]{Nalewajko2014}.

%On top of these parameter spaces we show the values $\Gamma$ and $r$ derived from the SED modeling. We can see in Figure~\ref{fig:nalewajko_constraints} that for TON~599 the modelling result is above the jet collimation limit.  

These parameter limits do not take into account the constraints given by the BLR and dust-torus opacity on the gamma-ray emission. We show with black vertical lines the minimum distance $r$ from the black hole in the SED models where the BLR would become fully opaque for the maximum observed energy $E_\mathrm{max}$ (370 GeV for PKS~1222+216 and 630 GeV for Ton~599). We consider the BLR opaque when $\tau_{\mathrm{BLR}, E_\mathrm{max}} > 5$, meaning that less than 1\% of the gamma rays can escape.
We can clearly see that considering the BLR opacity significantly tightens the constraints on the gamma-ray emission location in Ton~599, as mentioned in the previous section. The opacity constraint on PKS~1222+216 is weaker, as in that case the blob does not have to be as deep inside the BLR to reproduce the observed EIC emission.

%For the 10 flares of 3C 279, we could make some conservative assumptions about the BLR and torus. We need information on the optical emission during the flares to do so.

\section{Neutrino Emission During VHE Flares}\label{sec:discussion_neutrinos}

Luminous gamma-ray flares of FSRQs are potential sources of PeV-scale (${\sim}100~\mathrm{TeV}$ -- ${\sim}10~\mathrm{PeV}$) neutrino emission \citep[e.g.][]{Mannheim1993, Dermer2014, Kadler2016}. While the lack of point sources observed in IceCube data suggests that FSRQs are not the dominant population of neutrino sources, the possibility of neutrino emission during rare, bright flares has not been excluded \citep{Murase2016}. While \citet{Righi2020} have suggested that the bulk of the average neutrino emission from FSRQs occurs in the sub-EeV -- EeV energy range, their results do not exclude PeV-scale neutrino emission during outlier states. In the SED modeling of the VHE flares of PKS~1222+216 and Ton~599 presented here, a purely leptonic model gives an adequate representation of the data, and performing full hadronic modeling is beyond the scope of this work.

However, we can place analytic constraints on the potential PeV-scale neutrino flux produced during these events by considering a lepto-hadronic scenario in which synchrotron emission from secondary electrons produced by pion decay contributes a subdominant component to the second peak of the SED, similar to models used to describe the flaring emission of TXS~0506+056 coincident with the detection of a neutrino by IceCube \citep[e.g.][]{Keivani2018, Cerruti2019, Gao2019, Reimer2019}. In this section, we make use of the assumptions and methods of \citet{Gao2017}, particularly Appendix A of that work. All quantities in the following equations are in the comoving frame of the blob, unless explicitly noted with the superscript ``ob''.

We consider neutrinos produced by the $p\gamma$ interaction via the $\Delta^+$(1232) resonance with threshold energy $\epsilon_{p\gamma,\mathrm{th}} \sim 0.3$ GeV. The characteristic proton energy is $E_{p,\mathrm{char}} \sim E_\nu/K_\nu \sim 2~\mathrm{PeV}$, where $E_\nu = E_\nu^\mathrm{ob}(1+z)/\Gamma$ and $K_\nu \sim 0.05$ \citep{Murase2014}. Therefore, to check whether these sources could in principle support PeV neutrino emission, we first estimate the maximum energy to which protons can be accelerated in the source without escaping, following \citet{Hillas1984}, as

\begin{equation}
    E_{p,\mathrm{max}} = Z e \beta c B R,
\end{equation}

\noindent where the atomic number $Z = 1$ for protons, $e$ is the elementary charge, $\beta = v / c \sim 1$ for highly relativistic particles, $c$ is the speed of light, $B$ is the magnetic field in the source, and $R$ is the size of the source. Using the values in Table~\ref{tab:sed_model_params}, the maximum energy to which protons could have been accelerated in the gamma-ray emission regions for the flares of PKS~1222+216 and Ton~599 is $E_p \sim 500$ PeV, equivalent to an upper limit on the neutrino energy of $E_\nu^\mathrm{ob} \sim 400$ PeV, so PeV-scale neutrino emission is certainly feasible.

A limit on the neutrino flux can be imposed by considering a steady state within the time period of the flare in which the synchrotron luminosity of the secondary electrons equals the power injected by pion decay. Because the lifetime of the ultra-high-energy protons cooling by photo-pion production may be longer than that of the electrons, the resulting model can be considered to provide an upper limit on the neutrino luminosity. The steady-state proton energy density at $E_{p,\mathrm{char}}$ is given by \citep{Gao2017}:

\begin{equation}\label{eq:proton_energy_density}
    u_p(E_{p,\mathrm{char}}) = \frac{\alpha_\mathrm{fs} \epsilon_{p\gamma,\mathrm{th}}}{c \sigma_{p\gamma} K_{p\gamma}K_{\pi \rightarrow e}} \frac{m_p c^2}{E_{p,\mathrm{char}}} \frac{\nu F_{\nu,\mathrm{2}}^\mathrm{ob}}{\nu F_{\nu,\mathrm{t}}^\mathrm{ob}},
\end{equation}

\noindent where $\alpha_\mathrm{fs} = 4c/(3R)$ is the free-streaming escape rate, $\sigma_{p\gamma} = 5.0\times10^{-28}~\mathrm{cm}^2$ is the $p\gamma$ cross-section, $K_{p\gamma} \sim 0.2$ is the average inelasticity for the proton in the $p\gamma$ interaction, $K_{\pi \rightarrow e} \sim 1/8$ is the fraction of energy transferred to $e^\pm$ pairs from pion decay, $\nu F_{\nu,\mathrm{2}}^\mathrm{ob}$ is the observed flux due to synchrotron emission from the secondary electrons, and $\nu F_{\nu,\mathrm{t}}^\mathrm{ob}$ is the observed flux of the target photons of the $p\gamma$ interaction at $E_\mathrm{t} \sim \epsilon_{p\gamma,\mathrm{th}} m_p c^2 / E_{p,\mathrm{char}}$, which is directly constrained by the \textit{Swift}-XRT measurement at $E_\mathrm{t}\Gamma/(1+z) \sim 2~\mathrm{keV}$. We can estimate $\nu F_{\nu,\mathrm{2}}^\mathrm{ob}$ from the SED at the synchrotron peak frequency of the secondary electrons at

\begin{equation}
\begin{split}
    \nu_2^\mathrm{ob} & = \frac{c e (K_{p\gamma}K_{\pi \rightarrow e})^2}{2\pi (m_e c^2)^3}\frac{\Gamma}{1 + z} B E_{p,\mathrm{char}}\\[1ex]
    & \approx 10^{22}~\left(\frac{\Gamma}{23}\right) \left(\frac{B}{30 ~\mathrm{mG}}\right) \left(\frac{E_{p,\mathrm{char}}}{2~\mathrm{PeV}}\right)~\mathrm{Hz},
\end{split}
\end{equation}

\noindent for redshift $z \sim 0.5$. The corresponding power in protons can be estimated as

\begin{equation}
    L_p \sim \Gamma^2 u_p(E_{p,\mathrm{char}}) \alpha_\mathrm{esc} V(R),
\end{equation}

\noindent where for simplicity we assume the proton escape time $\alpha_\mathrm{esc} = 0.1\alpha_\mathrm{fs}$ and let $V(R) = 4\pi R^3/3$ for a spherical blob. Parameterizing this power by the Eddington luminosity boosted into the jet frame yields

\begin{equation}
\begin{split}
    L_p \sim 1 &\left(\frac{\Gamma}{23}\right)^2\left(\frac{R}{6\times10^{16}~\mathrm{cm}}\right) \left(\frac{M}{5\times10^{10}~\mathrm{M}_\odot}\right)^{-1}\\[1ex]
    \times &\left(\frac{E_{p,\mathrm{char}}}{2~\mathrm{PeV}}\right)^{-1} \left(\frac{\nu F_{\nu,2}^\mathrm{ob} / \nu F_{\nu,\mathrm{t}}^\mathrm{ob}}{0.5}\right)~L_\mathrm{Edd},
\end{split}
\end{equation}

\noindent where $\nu F_{\nu,2}^\mathrm{ob} \lesssim 0.5~\nu F_{\nu,\mathrm{t}}^\mathrm{ob} \sim 1\times10^{-12}~\mathrm{erg}~\mathrm{cm}^{-2}~\mathrm{s}^{-1}$ is a conservative estimate of the largest energetically reasonable contribution\footnote{This assumption requires about $5\times10^{-3}$ protons for every electron, from Eq.~\ref{eq:proton_energy_density} and the electron energy distributions reported in Table~\ref{tab:sed_model_params}.} to the SED at ${\sim}10^{22}$ Hz. The contribution is clearly subdominant. We can then estimate the observed neutrino energy flux, where $\dot{N}_{p\gamma}$ is the $p\gamma$ event rate per physical volume, using the relation

\begin{equation}
    \dot{N}_{p\gamma} \sim c\sigma_{p\gamma}\frac{u_\mathrm{ph}(E_\mathrm{t})}{E_\mathrm{t}}\frac{u_p(E_{p,\mathrm{char}})}{E_{p,\mathrm{char}}} = \frac{\alpha_\mathrm{fs}u_\nu(E_\nu)}{E_\nu},
\end{equation}

\noindent where $u_\mathrm{ph}(E_\mathrm{t})$ is the energy density of the target photons. Since $u_\nu(E_\nu)/u_\mathrm{ph}(E_\mathrm{t}) = \nu F_{\nu,\nu}^\mathrm{ob} / \nu F_{\nu,\mathrm{t}}^\mathrm{ob}$, we obtain the simple relation

\begin{equation}
    \nu F_{\nu,\nu}^\mathrm{ob} = \frac{K_\nu}{K_{p\gamma} K_{\pi\rightarrow e}} \nu F_{\nu,2}^\mathrm{ob} \sim 2~\nu F_{\nu,2}^\mathrm{ob}.
\end{equation}

The number of PeV-scale neutrinos of any flavor expected to be detected by IceCube during the VHE flare of PKS~1222+216 or Ton~599 is then

\begin{equation}
\begin{split}
    N_\nu \lesssim 0.001 &\left( \frac{\nu F_{\nu,\nu}^\mathrm{ob}}{2\times10^{-12}~\mathrm{erg}~\mathrm{cm}^{-2}~\mathrm{s}^{-1}}\right)\\[1ex]
    \times &\left(\frac{\Delta T}{5~\mathrm{day}}\right) \left(\frac{\mathcal{A}_\mathrm{eff}}{10^6~\mathrm{cm}^2}\right),
\end{split}
\end{equation}

\noindent where $\Delta T$ is the duration of the VHE flare, $\mathcal{A}_\mathrm{eff} \sim 10^6~\mathrm{cm}^2$ is the IceCube effective area for extremely high-energy real-time alerts in the PeV range \citep{Aartsen2017}, and $\Delta \nu \sim \ln(10)$ is assumed for the width of the neutrino spectrum. We conclude that it is plausible that PKS~1222+216 and Ton~599 could have produced PeV-scale neutrinos during their VHE flaring activity at a flux consistent with a null detection by current instruments.

To reduce the model-dependence of our constraints, $R$ could also be estimated using the timescale of gamma-ray flare variability,

\begin{equation}
    R \sim \frac{\delta}{1 + z} c \Delta T,
\end{equation}

\noindent from which estimates of $R \sim 7\times10^{17}~\mathrm{cm}~\mathrm{and}~1\times10^{17}~\mathrm{cm}$ are obtained for PKS~1222+216 and Ton~599. For the two sources, the constraints on the maximum neutrino energy are loosened to $E_\nu^\mathrm{ob} \sim 5~\mathrm{EeV}~\mathrm{and}~700~\mathrm{PeV}$, respectively, and the required proton luminosities are increased by a factor of ${\sim}10$ and ${\sim}2$, or within a few times the Eddington luminosity for both sources.

\section{Conclusions}\label{sec:conclusion}

In this paper, we present an analysis of the gamma-ray and multiwavelength emission of three bright, variable FSRQs observed by \textit{Fermi}-LAT and VERITAS: 3C~279, PKS~1222+216, and Ton~599, making use of almost 100 hours of VERITAS observations of these sources. No VHE gamma-ray activity was observed during multiple flares of 3C~279, which is the brightest of the three sources as observed with \textit{Fermi}-LAT, but VERITAS detected flares of both PKS~1222+216 and Ton~599.

The flux distributions of the \textit{Fermi}-LAT light curves of all three sources are consistent with the PDF derived from the stochastic differential equation proposed by \citet{Tavecchio2020}, in which the timescale of variability is controlled by processes in the accretion disk. The timescales associated with magneto-rotational instabilities and magnetic flux accumulation in the accretion disk obtained from this model are consistent with theoretical estimates.

We selected gamma-ray flaring states from the \textit{Fermi}-LAT light curves of the three sources using a procedure based on Bayesian blocks. Daily to sub-daily variability was observed by \textit{Fermi}-LAT during the flaring states. No pattern of asymmetry was found in the rise and decay times of the exponential components of the ten identified flares of 3C~279, while the fluence distribution of the flare components extended over one order of magnitude.

All three sources have similar gamma-ray spectra consistent with a log-parabola model. The average flaring spectrum of 3C~279 was found to exhibit stronger curvature than the baseline state, consistent with a null VHE detection even during extremely bright flares.

The SEDs of VHE flares of PKS~1222+216 and Ton~599 are described well by a purely leptonic emission model including an external inverse Compton emission component. For both sources, a strong constraint was placed on the Doppler factor, which must be ${\gtrsim}40$ for PKS~1222+216 and ${\gtrsim}50$ for Ton~599, to produce the observed gamma-ray emission at up to TeV energies despite internal absorption. We constrained the jet Lorentz factor and distance of the gamma-ray emission region from the central black hole using the independent method of \citet{Nalewajko2014}, which we augmented using an opacity constraint derived from our SED modeling. 
We found that both sources are operating in a similar regime, with both of them having strongly matter-dominated energetics and a gamma-ray emission zone located a few parsecs from the SMBH. It would be interesting in future work to extend this study to a large sample of TeV FSRQs in order to define the common physical properties of this blazar sub-class.

We calculated analytic constraints on a supposed subdominant hadronic component to the leptonic SED model, from which we estimated upper limits on potential PeV-scale neutrino emission during the TeV flares. We found that neutrino emission is energetically plausible at a flux consistent with a null detection by IceCube during the TeV flares.

%\textit{The observed flux excess at $\sim100$ MeV in the SED of TON~599 is still an open issue that could be addressed in further studies. While an EIC process over the dust torus radiation field would be the simplest possibility, we cannot exclude an hadronic origin (more details?) }

\acknowledgments

This research is supported by grants from the U.S. Department of Energy Office of Science, the U.S. National Science Foundation and the Smithsonian Institution, by NSERC in Canada, and by the Helmholtz Association in Germany. This research used resources provided by the Open Science Grid, which is supported by the National Science Foundation and the U.S. Department of Energy's Office of Science, and resources of the National Energy Research Scientific Computing Center (NERSC), a U.S. Department of Energy Office of Science User Facility operated under Contract No. DE-AC02-05CH11231. We acknowledge the excellent work of the technical support staff at the Fred Lawrence Whipple Observatory and at the collaborating institutions in the construction and operation of the instrument.

The \textit{Fermi} LAT Collaboration acknowledges generous ongoing support
from a number of agencies and institutes that have supported both the
development and the operation of the LAT as well as scientific data analysis.
These include the National Aeronautics and Space Administration and the
Department of Energy in the United States, the Commissariat \`a l'Energie Atomique
and the Centre National de la Recherche Scientifique / Institut National de Physique
Nucl\'eaire et de Physique des Particules in France, the Agenzia Spaziale Italiana
and the Istituto Nazionale di Fisica Nucleare in Italy, the Ministry of Education,
Culture, Sports, Science and Technology (MEXT), High Energy Accelerator Research
Organization (KEK) and Japan Aerospace Exploration Agency (JAXA) in Japan, and
the K.~A.~Wallenberg Foundation, the Swedish Research Council and the
Swedish National Space Board in Sweden.
 
Additional support for science analysis during the operations phase is gratefully
acknowledged from the Istituto Nazionale di Astrofisica in Italy and the Centre
National d'\'Etudes Spatiales in France. This work performed in part under DOE
Contract DE-AC02-76SF00515.

Data from the Steward Observatory spectropolarimetric monitoring project were used. This program is supported by Fermi Guest Investigator grants NNX08AW56G, NNX09AU10G, NNX12AO93G, and NNX15AU81G. L.S. acknowledges support from the Sloan Fellowship, the Cottrell Scholars Award, DoE DE-SC0016542, NASA 80NSSC18K1104 and NSF PHY-1903412.
This material is based upon work supported in part by the National Science Foundation under Grant PHY-1659528. % Meg Houck REU

% This research has made use of data from the OVRO 40-m monitoring program (Richards, J. L. et al. 2011, ApJS, 194, 29) which is supported in part by NASA grants NNX08AW31G, NNX11A043G, and NNX14AQ89G and NSF grants AST-0808050 and AST-1109911.

A.B.'s work was supported by the NSF Grant PHY-1806554 at Barnard College, Columbia University. O.H. thanks NSF for support under grants PHY-1707432 and PHY-2011420.
J.V. was partially supported by the Alliance program, a partnership between Columbia University, NY, USA, and three major French institutions: \'Ecole Polytechnique, Paris 1 Panth\'eon-Sorbonne University and Science Po. This material is based upon work supported by NASA under award number 80GSFC21M0002. S.B. acknowledges financial support by the European Research Council for the ERC Starting grant MessMapp, under contract no. 949555.

This work made use of the online cosmological calculator of \citet{Wright2006} and of NASA's Astrophysics Data System. We would like to thank the anonymous referee for the helpful and timely comments that helped improve the quality of the paper.

\software{
Astropy \citep{astropy:2013, astropy:2018},
Fermitools \citep{Fermitools2019},
NumPy \citep{numpy2020},
Matplotlib \citep{matplotlib2007},
SciPy \citep{scipy2020}.
}

\bibliography{bibliography}

\appendix

\section{Complete set of \textit{Fermi}-LAT flare profiles for 3C 279}\label{appendix:3c279flareprofiles}
\begin{figure*}[htb]
\centering
\includegraphics[width=\textwidth,height=0.8\textheight,keepaspectratio]{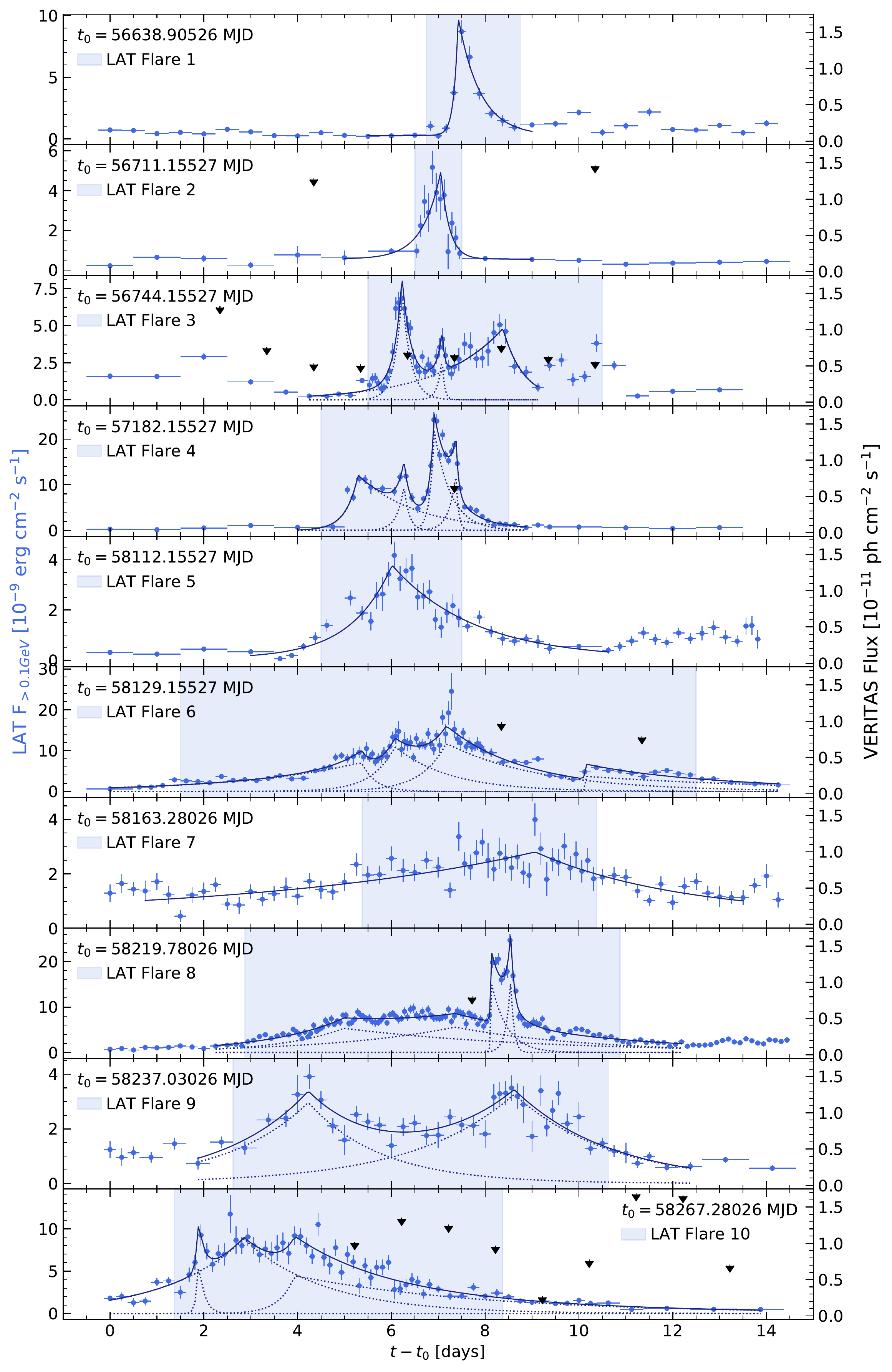}
\caption{3C 279 LAT sub-daily light curves (blue points) around the flaring episodes selected as described in Section \ref{sec:flarealgorithm} (shaded areas). The dotted blue lines show the fitted exponential profiles, with their sums shown in solid blue. The VERITAS 95\% upper limits are shown as black downwards arrows. For Flares 1, 5, 7, and 9, no VERITAS observations were taken around the time of the flare.}\label{fig:flareprofile3C279_all}
\end{figure*}

\begin{deluxetable*}{ccccc}
\tablecaption{Results of the LAT flare profile fits for 3C~279.\label{tab:flareprofile_3C279}}
%\tabletypesize{\scriptsize}
\tablehead{
\colhead{Amplitude ($F_0$)} & \colhead{$t_{\rm{peak}}$} &  \colhead{$t_{\rm{rise}}$} & \colhead{$t_{\rm{decay}}$}  & \colhead{Constant ($F_{\rm{const.}}$)}\\
\colhead{($10^{-9} $ erg cm$^{-2}\,$s$^{-1}$)}    & \colhead{(MJD)}  &  \colhead{(min.)}  & \colhead{(min.)} & \colhead{($10^{-9} $ erg cm$^{-2}\,$s$^{-1}$)}%\tablenotemark{a}}
}
\startdata
\multicolumn{5}{c}{Flare 1 (MJD 56645.655 -- 56647.655): $\chi^2/$d.o.f.= 12.05/8 = 1.51} \\\hline
9.56 $\pm$ 1.07 & 56646.330 $\pm$ 0.033 & 130 $\pm$ 45 & 674 $\pm$ 73 & 0.28 $\pm$ 0.06 \\\hline
\multicolumn{5}{c}{Flare 2 (MJD 56717.655 -- 56718.655): $\chi^2/$d.o.f.= 16.63/10 = 1.66} \\\hline
4.40 $\pm$ 0.70 & 56718.142 $\pm$ 0.043 & 445 $\pm$ 95 & 307 $\pm$ 86 & 0.55 $\pm$ 0.07 \\\hline
\multicolumn{5}{c}{Flare 3 (MJD 56749.655 -- 56754.655): $\chi^2/$d.o.f.= 69.98/34 = 2.06} \\\hline
7.27 $\pm$ 0.64 & 56750.382 $\pm$ 0.015 & 229 $\pm$ 25 & 267 $\pm$ 42 & N.A. \\
2.78 $\pm$ 1.15 & 56751.238 $\pm$ 0.024 & 140 $\pm$ 82 & 69 $\pm$ 47 & N.A. \\
4.80 $\pm$ 0.37 & 56752.532 $\pm$ 0.067 & 2001 $\pm$ 116 & 631 $\pm$ 136 & N.A. \\\hline
\multicolumn{5}{c}{Flare 4 (MJD 57186.655 -- 57190.655): $\chi^2/$d.o.f.= 77.31/19 = 4.07} \\\hline
12.07 $\pm$ 0.67 & 57187.446 $\pm$ 0.031 & 378 $\pm$ 46 & 1784 $\pm$ 147 & N.A. \\
\phn9.79 $\pm$ 2.29 & 57188.425 $\pm$ 0.028 & 216 $\pm$ 101 & 155 $\pm$ 64 & N.A. \\
21.72 $\pm$ 1.59 & 57189.069 $\pm$ 0.008 & 137 $\pm$ 18 & 512 $\pm$ 55 & N.A. \\
12.41 $\pm$ 1.30 & 57189.532 $\pm$ 0.010 & 220 $\pm$ 63 & 77 $\pm$ 25 & N.A. \\\hline
\multicolumn{5}{c}{Flare 5 (MJD 58116.655 -- 58119.655): $\chi^2/$d.o.f.= 54.70/29 = 1.89} \\\hline
\phn3.72 $\pm$ 0.20 & 58118.171 $\pm$ 0.069 & 1278 $\pm$ 220 & 2521 $\pm$ 309 & 0.06 $\pm$ 0.11 \\\hline
\multicolumn{5}{c}{Flare 6 (MJD 58130.655 -- 58141.655): $\chi^2/$d.o.f.= 141.28/72 = 1.96} \\\hline
\phn7.08 $\pm$ 1.01 & 58134.520 $\pm$ 0.055 & 3719 $\pm$ 390 & 421 $\pm$ 259 & N.A. \\
10.95 $\pm$ 3.74 & 58135.229 $\pm$ 0.053 & 718 $\pm$ 232 & 3535 $\pm$ 1394 & N.A. \\
11.78 $\pm$ 7.00 & 58136.266 $\pm$ 0.048 & 349 $\pm$ 160 & 1839 $\pm$ 1055 & N.A. \\
\phn3.44 $\pm$ 0.66 & 58139.546 $\pm$ 0.033 & 233 $\pm$ 175 & 6119 $\pm$ 1824 & N.A. \\\hline
\multicolumn{5}{c}{Flare 7 (MJD 58168.655 -- 58173.655): $\chi^2/$d.o.f.= 78.81/58 = 1.36} \\\hline
\phn2.36 $\pm$ 0.50 & 58172.345 $\pm$ 0.242 & 8540 $\pm$ 4159 & 4458 $\pm$ 2319 & 0.45 $\pm$ 0.59 \\\hline
\multicolumn{5}{c}{Flare 8 (MJD 58222.655 -- 58230.655): $\chi^2/$d.o.f.= 177.25/106 = 1.67} \\\hline
\phn5.29 $\pm$ 1.29 & 58224.773 $\pm$ 0.105 & 1996 $\pm$ 716 & 5899 $\pm$ 4035 & N.A. \\
17.70 $\pm$ 2.01 & 58227.945 $\pm$ 0.004 & \bf{36 $\mathbf{\pm}$ 13} & 329 $\pm$ 131 & N.A. \\
16.42 $\pm$ 1.87 & 58228.323 $\pm$ 0.012 & 140 $\pm$ 54 & 115 $\pm$ 48 & N.A. \\
\phn5.59 $\pm$ 1.69 & 58227.139 $\pm$ 0.133 & 3816 $\pm$ 1450 & 4077 $\pm$ 2080 & N.A. \\\hline
\multicolumn{5}{c}{Flare 9 (MJD 58239.655 -- 58247.655): $\chi^2/$d.o.f.= 46.25/34 = 1.36} \\\hline
\phn2.96 $\pm$ 0.40 & 58241.258 $\pm$ 0.149 & 2546 $\pm$ 595 & 2226 $\pm$ 1088 & N.A. \\
\phn3.27 $\pm$ 0.25 & 58245.648 $\pm$ 0.133 & 3080 $\pm$ 1384 & 3028 $\pm$ 303 & N.A. \\\hline
\multicolumn{5}{c}{Flare 10 (MJD 58268.655 -- 58275.655): $\chi^2/$d.o.f.= 75.80/55 = 1.37} \\\hline
\phn6.23 $\pm$ 9.46 & 58269.171 $\pm$ 0.182 & 73 $\pm$ 236 & 177 $\pm$ 102 & N.A. \\
\phn8.81 $\pm$ 0.84 & 58270.137 $\pm$ 0.107 & 2392 $\pm$ 243 & 2449 $\pm$ 956 & N.A. \\
\phn4.46 $\pm$ 1.91 & 58271.223 $\pm$ 0.088 & 477 $\pm$ 431 & 5824 $\pm$ 862 & N.A. \\
\enddata
\tablecomments{The smallest variability time found is indicated in boldface.}
\end{deluxetable*}

\section{\textit{Fermi}-LAT spectral fit parameters}\label{appendix:spectral_fit_parameters}

\begin{deluxetable*}{ccccccc}[htbp]
\tablecaption{\textit{Fermi}-LAT spectral fit parameters.
\label{tab:flare_fitpar}}
\tablehead{\colhead{State} & \colhead{Epoch} & \colhead{TS} & \colhead{$N_0$} & \colhead{$\alpha$} & \colhead{$\beta$} & \colhead{Flux} \\
 & \colhead{(MJD)} & \colhead{} & \colhead{($10^{-10}\,$MeV$^{-1}\,$cm$^{-2}\,$s$^{-1}$)} &  & \colhead{($\times\,10^{-2}$)} & \colhead{($10^{-6}\,$ph$\,$cm$^{-2}\,$s$^{-1}$)}}
\startdata
\multicolumn{7}{c}{\textbf{3C 279}} \\\hline
Global & 54682.66 -- 58459.35 & 271945 & 3.33 $\pm$ 0.02 & 2.228 $\pm$ 0.004 & 6.1 $\pm$ 0.3 & 0.751 $\pm$ 0.004 \\
Low state & 56230.66 -- 56465.66 & 1130 & 0.54 $\pm$ 0.03 & 2.38 $\pm$ 0.06 & 2.9 $\pm$ 3.1 & 0.14 $\pm$ 0.01 \\
VER-LAT quiescent & Various & 322 & 5.7 $\pm$ 0.8 & 2.2 $\pm$ 0.1 & 1.8 $\pm$ 7.0 & 1.4 $\pm$ 0.2 \\\hline
\multirow{4}{1.7cm}{LAT Flares simultaneous with VER obs.} & 56750.27 -- 56750.34 & 578 & 69 $\pm$ 10 & 2.1 $\pm$ 0.1 & 31 $\pm$ 13 & 11 $\pm$ 2 \\
 & 57189.17 -- 57189.23 & 1141 & 173 $\pm$ 15\phn & 2.07 $\pm$ 0.07 & 14 $\pm$ 6 & 32 $\pm$ 3 \\
 & 58227.22 -- 58227.27 & 355 & 79 $\pm$ 14 & 1.8 $\pm$ 0.2 & 21 $\pm$ 13 & 12 $\pm$ 4 \\
 & 58272.18 -- 58272.22 & 235 & 62 $\pm$ 15 & 1.5 $\pm$ 0.3 & 42 $\pm$ 2 & 7 $\pm$ 3 \\\hline
\multirow{10}{1.7cm}{LAT Flares} & 56645.66 -- 56647.66 & 1633 & 22.0 $\pm$ 1.5 & 1.73 $\pm$ 0.07 & 9.6 $\pm$ 3.2 & 3.7 $\pm$ 0.3 \\
 & 56717.66 -- 56718.66 & 900 & 23.2 $\pm$ 2.0 & 2.08 $\pm$ 0.08 & 11 $\pm$ 6 & 4.5 $\pm$ 0.5 \\
 & 56749.66 -- 56754.66 & 7680 & 24.0 $\pm$ 0.8 & 2.20 $\pm$ 0.03 & 13 $\pm$ 2 & 5.0 $\pm$ 0.2 \\
 & 57186.66 -- 57190.66 & 23623 & 77.9 $\pm$ 1.5 & 2.04 $\pm$ 0.02 & 11 $\pm$ 1 & 14.7 $\pm$ 0.5 \\
 & 58116.66 -- 58119.66 & 3543 & 19.5 $\pm$ 0.9 & 2.06 $\pm$ 0.04 & 4.1 $\pm$ 2.2 & 4.0 $\pm$ 0.2 \\
 & 58130.66 -- 58141.66 & 27256 & 53.3 $\pm$ 0.9 & 2.14 $\pm$ 0.02 & 8.5 $\pm$ 1.1 & 11.0 $\pm$ 0.2 \\
 & 58168.66 -- 58173.66 & 4932 & 19.5 $\pm$ 0.7 & 2.10 $\pm$ 0.03 & 5.8 $\pm$ 2.0 & 4.1 $\pm$ 0.2 \\
 & 58222.66 -- 58230.66 & 53745 & 59.4 $\pm$ 0.7 & 2.00 $\pm$ 0.01 & 9.6 $\pm$ 0.7 & 11.2 $\pm$ 0.1 \\
 & 58239.66 -- 58247.66 & 7989 & 19.2 $\pm$ 0.6 & 1.90 $\pm$ 0.03 & 6.2 $\pm$ 1.4 & 3.6 $\pm$ 0.1 \\
 & 58268.66 -- 58275.66 & 107456 & 47.8 $\pm$ 1.3 & 1.91 $\pm$ 0.02 & 14 $\pm$ 2 & 8.2 $\pm$ 0.2 \\\hline
\multicolumn{7}{c}{\textbf{PKS 1222+216}} \\\hline
Global & 54682.66 -- 58459.64 & 94556 & 1.66 $\pm$ 0.01 & 2.305 $\pm$ 0.007 & 3.8 $\pm$ 0.4 & 0.337 $\pm$ 0.002 \\\hline
\multicolumn{7}{c}{\textbf{Ton 599}} \\\hline
Global & 54682.66 -- 58464.49 & 48176 & 6.55 $\pm$ 0.06 & 2.11 $\pm$ 0.01 & 5.5 $\pm$ 0.5 & 0.161 $\pm$ 0.002 \\
\enddata
\end{deluxetable*}

\end{document}